\documentclass[aps,prd,onecolumn,nofootinbib,preprintnumbers,superscriptaddress]{revtex4}
\usepackage[mathscr]{eucal}
\usepackage{mathtools}
\usepackage{color}
\usepackage{graphicx}
\usepackage{epsf}
\usepackage{dcolumn}
\usepackage{bm}
\usepackage{amsmath}
\usepackage{amssymb}
\usepackage{amsfonts}
\usepackage{amsthm}
\usepackage{amscd}

\def\slashchar#1{\setbox0=\hbox{$#1$}     		
   \dimen0=\wd0                                 	
   \setbox1=\hbox{/} \dimen1=\wd1               	
   \ifdim\dimen0>\dimen1                        	
      \rlap{\hbox to \dimen0{\hfil/\hfil}}      	
      #1                                        	
   \else                                        	
      \rlap{\hbox to \dimen1{\hfil$#1$\hfil}}   	
      /                                         	
   \fi}

\begin{document}

\preprint{ITEP-TH-21/17}

\title{On S-duality for holographic p-wave superconductors}

\author{Alexander Gorsky}
\affiliation{Institute for Information Transmission Problems, B.Karetnyi 19, Moscow, Russia}
\affiliation{Moscow Institute of Physics and Technology, Dolgoprudny 141700, Russia}
\author{Elena Gubankova}
\affiliation{Department of Mathematics and Statistics, Boston University, Boston, MA 02215, USA}
\affiliation{Institute for Theoretical and Experimental Physics, B. Cheryomushkinskaya 25, Moscow 117218, Russia}
\affiliation{Skolkovo Institute of Science and Technology, Skolkovo Innovation Center, Moscow 143026, Russia}
\author{Ren\'e Meyer}
\affiliation{Institute for Theoretical Physics and Astrophysics, University of W\"urzburg, 97074 W\"urzburg, Germany}
\author{Andrey Zayakin}
\affiliation{Institute for Theoretical and Experimental Physics, B. Cheryomushkinskaya 25, Moscow 117218, Russia}

\begin{abstract}
We consider the generalization of the S-duality transformation previously
investigated in the context of the FQHE and s-wave superconductivity to p-wave
superconductivity in 2+1 dimensions in the framework of the AdS/CFT correspondence. The vector Cooper condensate transforms under the 
S-duality action to the pseudovector condensate at the dual side. The 3+1-dimensional Einstein-Yang-Mills theory,  the holographic dual to p-wave superconductivity, is used to investigate the S-duality action via the AdS/CFT correspondence. 
It is shown that in order to implement the duality transformation, chemical potentials both on the electric and magnetic side of the duality have to be introduced. 
A relation for the product of the nonabelian conductivities in the dual models is derived. 
We also conjecture a flavor S-duality transformation in the holographic dual to 3+1-dimensional QCD low-energy QCD with non-abelian flavor gauge groups. The conjectured S-duality interchanges isospin and baryonic chemical potentials.

Keywords: AdS/CFT correspondence, S-duality, particle-vortex duality, BCS, Gross-Neveu, magnetic catalysis
\end{abstract}

\maketitle

\section{Introduction}\label{intro}

The AdS/CFT correspondence opens a new route to studying strongly correlated systems with bosonic and fermionic degrees of freedoms at finite chemical potential and density. 
In gauge/gravity duality, the charge density $\rho$ conjugate to the chemical potential is dual to an electric flux
emanating from the boundary of a space-time which is asymptotically Anti-de Sitter. Due to flux conservation, this charge has to reside behind the horizon of a black hole in the interior of the AdS space-time. A magnetic field in the dual field theory corresponds to switching on a magnetic flux component in the bulk space-time. 

The S-duality transformation, which exchanges electric with magnetic field strengths, is a well-studied symmetry for the U(1) Einstein-Maxwell theory in 3+1 dimensional asymptotically AdS space-times. 
It acts non-trivially on the boundary conditions of the U(1) gauge fields, exchanging Neumann to Dirichlet boundary conditions \cite{Witten}, in this way exchanging electrically charged with magnetically charged black holes, and hence charge density with magnetic field in the dual field theory. 
It corresponds to particle-vortex duality in the 2+1-dimensional dual field theory, and also acts in this way on the conductivities in the field theory. 

S-duality together with the T-duality transformation generates the group of modular transformations $SL(2,Z)$.  The bulk S duality acts naturally on the conserved U(1) currents of 2+1-dimensional conformal field theories \cite{Witten}, and corresponds to 3d mirror symmetry \cite{kapustin}. 
The most studied example of subgroups of the $SL(2,Z)$ symmetry in 2+1 dimensions is in the framework of the FQHE. It was observed long ago \cite{Burgess} that
a combination of the conductivities $\sigma = \sigma_{xy}+i \sigma_{xx}$ transforms fractionally linear under   modular transformations, as does the filling fraction $\nu = \frac{\rho}{B}$, where $\rho$ is density and $B$ is the magnetic field. The modular subgroups $\Gamma_0(2)\subset SL(2,Z)$ and $\Gamma_\theta(2)\subset SL(2,Z)$ act as flux attachment transformations on, respectively, the odd-denominator and even-denominator filling fractions tates. The full $SL(2,Z)$ was argued to be relevant for certain ${\cal N}=2$ supersymmetric field theories \cite{dolan:2006zc}. The generator of the S-duality transformation 
$\nu \rightarrow \frac{1}{\nu}$ interchanges the density and magnetic field and can be understood as particle-hole transformation \cite{Witten}. The nontrivial modular properties of the complex conductivity $\sigma = \sigma_{xy}+i \sigma_{xx}$ can be argued to hold along the RG 
 flow for energies sufficiently low such that higher derivative operators in the external electromagnetic field can be neglected \cite{Burgess} and the RG flow itself happens on the two-dimensional submanifold in coupling space parametrized by $\sigma$.  The $SL(2,Z)$ (subgroup) action 
in the FQHE was implemented as the S-duality transformation in several 3+1-dimensional holographic dual systems  \cite{burg2, karch, rene},  
where the dual gravitational theory used was an Einstein-Maxwell-dilaton-axion 
model. The state parameters $\rho, B$ are mapped to the charge and magnetic field emanating from the bulk black hole solution, and the 
modular transformation of $\sigma_{xy} +i \sigma_{xx}$ is related to the transformation of the bulk axio-dilaton field and the 
boundary two-point correlators for the conductivities via the Kubo formula, where the axio-dilaton encodes  coupling constants $\sigma_{xy}$ and $\sigma_{xx}$ governing the two-dimensional low energy RG flow.

Having understood the holographic S-duality action in the simplest case of a single conserved $U(1)$ current in the presence of charge density and magnetc field, the next natural step is to allow for charged scalar operators to condense in the context of s-wave holographic superconductivity\footnote{In most holographic systems the $U(1)$ symmetry which gets broken in the superconducting phase is actually a global $U(1)$ which was ungauged during the process of holographic renormalization. During this process, the UV cutoff is taken to infinity, and gauge symmetries become global. Nevertheless, the physics of the broken state more closely resembles a superconductor than a superfluid, as it e.g. admits a dynamically generated gap.} in the $(\mu, B)$ space of the grand-canonical ensemble. The new feature introduced by superconductivity is the presence of the condensate and the mass gap and, since holographic superconductors are type II, Abrikosov vortices in the presence of an external magnetic field. The natural guess for the duality mapping of the S-duality action, which exchanges finite density and zero magnetic field with finite magnetic field and zero density, is  
the that the superconducting state with the charged order parameter condensate is mapped to the the magnetic catalysis problem triggered by the finite magnetic field and which admits a neutral  excitonic condensate. 
%
Indeed such a mapping between the BCS and magnetic catalysis (MC) models has been discussed in (1+1) dimensions (here
DOS stands for the density of states, $h$ is the magnetic field strength and $\mu$ is the chemical potential) \cite{Schmitt}:

\begin{table}[h]
\begin{center}
\begin{tabular}{|c|c|}
\hline 
${\rm MC}$ & ${\rm BCS}$\\
\hline 
$(3+1){\rm d}\to (1+1){\rm d}\;{\rm in}\;x-{\rm space}$ & $(1+1){\rm d}\;{\rm in}\; p-{\rm space}$\\ 
${\rm LLL\;\; and}\;\; \varepsilon=0\;\;{\rm surface}$ & ${\rm Fermi\;\;surface}\;\; \varepsilon=\mu$ \\
$\varepsilon=\sqrt{p_z^2+2|eh|n}$ & $\varepsilon=p-p_F,\; p=\sqrt{\vec{p}^2}$ \\
${\rm excitonic:}\;\; \Delta\sim G\langle\bar{\psi}\psi\rangle$ & ${\rm SC:}\;\;\Delta\sim G\langle\psi\psi\rangle$ \\
$\Delta\sim \sqrt{eh}\;{\rm exp}(-\frac{const}{G\nu_0})$ &
$\Delta\sim \mu\;{\rm exp}(-\frac{const}{G\nu_F})$\\
$\nu_0\;\; {\rm is\;\; DOS\;\;at}\;\; \varepsilon=0$ & $\nu_F \;\;{\rm is \;\;DOS\;\; at}\;\;\varepsilon=\mu$ \\
$h\;\;{\rm enhances},\;\;\mu\;\;{\rm destroys}\;\;\Delta$ & $\mu\;\;{\rm enhances},\;\;h\;\;{\rm destroys}\;\;\Delta$ \\
$\delta\Omega\sim h(\mu^2-\frac{\Delta^2}{2})$ & $\delta\Omega\sim \mu^2(\delta\mu^2 -\frac{\Delta^2}{2})$\\
$h\gg\mu,\Delta$ & $\mu\gg\delta\mu,\Delta$ \\
${\rm it\;\;can\;\;have}\;\;\mu=0$ & ${\rm it\;\; can\;\;have}\;\;h=0$ \\
$T_c \;\; {\rm grows\;\; with}\;\;h\;\;({\rm MC})$ & $T_c\;\;{\rm decreases\;\;with}\;\; h$\\
$T_c\;\;{\rm decreases\;\;with}\;\;\mu$ & $T_c\;\;{\rm grows\;\;with}\;\; \mu\;\;({\rm SC})$\\
\hline 
\end{tabular}
\caption{Mapping between the BCS and magnetic catalysis models.} 
\label{mapping}
\end{center}
\end{table} 
Effectively one dimensional dynamics in both cases leads to similarities in formulas for the pairing gap $\Delta$
and the gain in thermodynamic potential $\delta\Omega$ as compared to the normal unpaired state. 
The parameters in the two systems at a nonzero density and at a nonzero magnetic field are mapped onto each other as follows: 
\begin{table}[h]
\begin{center}
\begin{tabular}{ccc}
${\rm MC}$ & $\longleftrightarrow$ & ${\rm BCS}$\\
$\langle\bar{\psi}\psi\rangle\neq 0$ & $\longleftrightarrow$ &     $\langle\psi\psi\rangle\neq 0$\\
${\rm finite}\;\; h$ & $\longleftrightarrow$ & ${\rm finite}\;\; \rho$ \\
${\rm small}\;\; \mu$ & $\longleftrightarrow$ & ${\rm small}\;\; \delta\mu$ \\
$h\gg\mu$ & $\longleftrightarrow$ & $\mu\gg\delta\mu$
\end{tabular}
\label{mapping2qcd}
\end{center}
\end{table}
The last line expresses a hierarchy of scales. A similar mapping has been obtained in case of the Gross-Neveu and the BCS model \cite{Thies}, where the
magnetic field $h$ maps to the chemical potential mismatch
$\delta\mu$ and is relevant for the inhomogeneous superconductors in 
the incommensurate phase. In the Sakai-Sugimoto model of holographic QCD \cite{Schmitt}, analytical formula
for the free energy difference between condensed and normal states have been obtained, proving the stability of both condensed states. 
In the strong magnetic field regime ("direct" magnetic catalysis), the free energy difference takes a remarkably simple
form \cite{Schmitt}
\begin{equation}
\delta\Omega\sim -h\left(\frac{\Delta(h)^2}{2}-\mu^2\right),
\label{Clogston-stability}
\end{equation}
which exactly maps to the result obtained in the field theory Table (\ref{mapping}).

In the gravity dual description the duality mapping between the two setups, superconductivity and magnetic catalysis (MC), is as follows: 
\begin{table}[h]
\begin{center}
\begin{tabular}{|c|c|}
\hline 
${\rm Holographic\;\;MC}$ & ${\rm Holographic\;\;SC}$\\
\hline 
${\rm dyonic\;\;AdS\;\;RN\;\; BH,}$ & ${\rm AdS\;\;RN\;\;BH}$\\
${\rm Schwarzschild\;\;BH}$ & \\
$|H|> |Q|$ & $|Q|>|H|$ \\
${\rm it\;\;can\;\;be}\;\;Q=0$ &${\rm it\;\;can\;\;be}\;\;H=0$ \\
$Z_2 ({\rm chiral\;\;SB})\;\;{\rm broken}$  & $U(1)\;\;{\rm broken}$\\
${\rm magnetic\;\;field\;\;enhances\;\;it}$ & ${\rm magnetic\;\;field\;\;destroys\;\;it}$\\
${\rm electric\;\;field\;\;destroys\;\;it}$ & ${\rm electric\;\;field\;\;enhances\;\;it}$\\
\hline
${\rm Callan}-{\rm Rubakov\;\;effect}$ & ${\rm dual\;\;Callan}-{\rm Rubakov\;\;effect}$\\
\hline 
\end{tabular}
\caption{Mapping between the BCS and magnetic catalysis models in the gravity dual description.} 
\label{duality}
\end{center}
\end{table}
Some explicit examples for the S-duality transformation between the 2+1 dimensional field theories at finite density are known in the literature: In \cite{Sachdev}, the mapping  between the XY model and the abelian Higgs model was studied. 
The superfluid phase corresponds to the condensate of the scalar while 
the solid phase is generated via the condensate of the monopole operator \cite{Sachdev}.
The solid state is a kind of Abrikosov lattice of vortices. Similar considerations 
have been made in the context of the $CP_1$ model \cite{SeibergSenthilWangWitten}, in which the form of the action is identical for the S-dual theory 
however the global U(1) and topological U(1) currents are interchanged . 
Some analysis of the corresponding holographic model which involves the electrically 
and magnetically charged black holes can be found in \cite{Sachdev}. The instabilities which yield the bulk
condensates were identified. In particular the cristalline phase at the boundary
corresponds to the  nontrivial magnetic condensate in the bulk. 
Another example concerns the mapping between the BCS model at large chemical potentials and 
Gross-Neveu model in strong external magnetic field \cite{Gubankova}. 

In this paper we go one step further and consider the S-duality action on a holographic p-wave superconductor. 
In a p-wave superconductor the order parameter is vector-like since Cooper pairing 
occurs in the  $L=1$ state.
There are two 
different holographic models of  p-wave superconductivity \cite{Gubser,cai}: a model 
with SU(2) gauge group \cite{Gubser} and an abelian bulk model with additional vector mesons \cite{cai}. 
We shall discuss the bulk SU(2) approach \cite{Gubser} and search for the 
S-duality action in the bulk theory and what it implies in the dual 2+1-dimensional boundary theory. 
We shall argue that looking
at the particular solutions to the bulk equation of motion for the condensate and the U(1) part of the SU(2) gauge field in the background of the charged BH in $AdS_4$, 
some general claims concerning the S-duality for the p-wave superconductor can be made. 
In particular we will show that in order to define the S-duality a dual
"`magnetic"'chemical potential has to be introduced. It will also be demonstrated that 
there is a relation between the conductivities in the S-dual theories similar to the one 
found in holographic models with abelian bulk fields \cite{burg2, karch, rene}. The 
vector order parameter is mapped onto the pseudovector of the dual SU(2) field strength under the S-duality transform.

In low-energy QCD the analogue of p-wave superconductivity occurs at 
non-vanishing isotopic chemical potential \cite{vectormes1,vectormes2,aharony,kaminski} where the vector mesons 
condense. Holographic QCD models like \cite{SakaiSugimoto} include a non-abelian 
flavor gauge theory in the dual 4+1 bulk geometry, and hence it is natural to ask whether it is possible to 
define a kind of S-duality transform in the dual bulk theory which then induces a S-duality action 
in the dual low-energy QCD at the boundary. To this aim we conjecture 
that the proper S-dual pair in 4+1 bulk are the isotopic U(1) and the topological U(1) 
symmetries generated by the topologically conserved current. At the 3+1 boundary this pair 
corresponds to the isotopic  and baryonic global charges. Therefore it is natural 
to conjecture that the flavor S-duality interchanges the low-energy QCD 
with isotopic and baryonic chemical potentials correspondingly. We give several consistency arguments in favor of this conjecture.

The paper is organized as follows. In section 2 we consider the holographic p-wave superconductor and 
construct its S-dual in the dual asymptotically $AdS_4$ space-time. 
In section 3 we solve the equations of motion numerically and check that the solutions satisfy 
the duality relations close to the phase transition. In section 4 we speculate on a possible S-duality relation for low-energy QCD with isotopic and baryonic chemical potentials. We conclude and discuss our findings in section 5. In the appendices
we prove $SL(2,Z)$ invariance of the SU(2) symmetric Axio-Dilaton-Yang-Mills action, and review the p-wave superconductor in the five-dimensional setting.

\section{P-wave superconductor and its S-dual}\label{sec:2}
\subsection{Equations of motion}
We base our consideration of the p-wave superconductor on the electric-magnetic duality for the non-abelian gauge fields 
in four dimensions.
In full analogy to U(1) gauge field case, we show the SL(2,Z) invariance of the Einstein-Maxwell action 
with the SU(2) gauge fields which are coupled to an axion and a dilaton fields in Appendix A.
The gauge-gravity action coupled to an axio-dilaton is 
\begin{eqnarray}
S_{\phi,\chi} &=& -\int d^4x \sqrt{-g}\left(\frac{1}{2\kappa^2}
\left[R-2\Lambda +\frac{1}{2}(\partial_{\mu}\phi\partial^{\mu}\phi
+{\rm e}^{2\phi}\partial_{\mu}\chi\partial^{\mu}\chi)
\right]\right.\nonumber\\
  && \left.+\frac{1}{4}{\rm e}^{-\phi}F_{\mu\nu}F^{\mu\nu} - \frac{1}{4}\chi F_{\mu\nu}\ast F^{\mu\nu}
\right)
\label{action}
\end{eqnarray}
where the scalar fields are dilaton $\phi$ and axion $\chi$, the field strength 
$F_{\mu\nu}=\partial_{\mu}A_{\nu}-\partial_{\nu}A_{\mu}+[A_{\mu},A_{\nu}]$ with SU(2) gauge field
$A_{\mu}=A_{\mu}^a \tau_a$. 
The dual field strength is obtained by applying the Hodge star operation  
$\ast F_{\mu\nu}:=\frac{1}{2}\epsilon_{\mu\nu\lambda\rho}F^{\lambda\rho}$,
where completely antisymmetric Levi-Civita tensor $\epsilon_{\mu\nu\rho\lambda}$ 
has a factor of $\sqrt{-g}$ with $g={\rm det}g_{\mu\nu}$ extracted and transforms as a tensor and not as a tensor density,
and indexes are freely raised and lowered using the metric $g_{\mu\nu}$ whose signature is Lorentzian
$(-\;+\;+\;+)$. 
The constant $\Lambda=3/L^2$ is the AdS cosmological constant
and $\kappa^2=8\pi G$ is the Newton's constant. The weak curvature means $\kappa^2/L^2 \ll 1$.
The relation to the gauge coupling $g$ and the theta-angle $\theta$ is 
\begin{equation}
{\rm e}^{-\phi}=\frac{1}{g_E^2}=\frac{1}{g^2},\;\;\;
\chi=\frac{1}{g_B^2}=\theta,
\end{equation}
where subscripts $E$ and $B$ stand for the electric and magnetic part.
Therefore weak coupling corresponds to ${\rm e}^{\phi}\ll 1$.
The $S$-operator acts on the axio-dilaton as
\begin{equation}
\tau\rightarrow \tilde{\tau} = -\frac{1}{\tau},
\label{Sdual_tau1}
\end{equation}
We restrict ourselves to the vanishing axion field
\begin{equation}
\chi=0.
\end{equation} 
According to eq.(\ref{Sdual_tau1}) and definition of $\tau=\chi+i{\rm e}^{-\phi}$, 
the axion field is not generated by the S-duality transformation.
In this case the S-operator acts on the gauge field strength as
\begin{equation}
F_{\mu\nu}^a\rightarrow \tilde{F}_{\mu\nu}^a = -{\rm e}^{-\phi}\ast F_{\mu\nu}^a.
\label{Sdual_F1}
\end{equation}
Eqs.(\ref{Sdual_tau1},\ref{Sdual_F1}) express a familiar electric-magnetic duality
where the field strength transforms into a Hodge-dual one and the coupling transformation is $g^2\rightarrow \frac{1}{g^2}$, 
therefore the weak-strong coupling regimes are interchanged. Electric-magnetic duality exists only in $(3+1)$ dimensions
where a two form $F_{\mu\nu}$ is dual to a two form again $\star F_{\mu\nu}$, 
as opposed for example to $(4+1)$-d where a two form is dual to a three form. In $(3+1)$-d the Hodge-dual is defined
as $\star F = \frac{\sqrt{-g}}{4}\epsilon_{\mu\nu\rho\sigma}F^{\rho\sigma}dx^{\mu}\wedge dx^{\nu}$. 

We start with the holographic p-wave superconductor introduced by Gubser et.al. \cite{Gubser}. 
It constitutes the electric side (E-side) of the duality. We will work in the four dimensional space,
in the background of Reissner-Nordstrom black hole with asymptotic $AdS_4$ in the UV, but for now 
we write a general metric 
\begin{equation}
ds^2 = -g_{tt}(z)dt^2 + g_{zz}(z)dz^2 + g_{xx}(z)dx^2 +g_{yy}(z) dy^2,
\label{metric}
\end{equation}
without off-diagonal terms because we restrict ourselves to the probe limit.
The non-abelian SU(2) gauge field has the following components \cite{Gubser}
\begin{equation}
A(z)= A_t^3(z)\tau^3 dt + A_x^1(z)\tau^1dx,
\label{p-wave}
\end{equation}
where $A_t^3$ plays the role of the chemical potential and $A_x^1$ is the component which condenses at $T_c$ and 
leads to spontaneous symmetry breaking and superconductivity.
The field strengths obtained with these two components of vector potential eq.(\ref{p-wave}) is
\begin{eqnarray}
&& F_{zt}^3 = \partial_z A_t^{3},\label{Fzt}\\
&& F_{zx}^1 = \partial_z A_x^{1},\label{Fzx}\\
&& F_{tx}^2 = A_t^3 A_x^1,\label{Ftx}
\end{eqnarray}
and we don't write other components such as $F_{zt}^3=-F_{tz}^3$.
Performing the S-duality transformation eqs.(\ref{Sdual_tau1},\ref{Sdual_F1}), we obtain the magnetic side (B-side) of the duality,
\begin{eqnarray}
&& \tilde{F}_{xy}^3 = -\frac{1}{g^2} \frac{\sqrt{-g}} {g_{tt}g_{zz}}F_{zt}^3, \label{tildeFxy}\\
&& \tilde{F}_{ty}^1 = -\frac{1}{g^2} \frac{\sqrt{-g}} {g_{zz}g_{xx}}F_{zx}^1, \label{tildeFty}\\
&& \tilde{F}_{zy}^2 = -\frac{1}{g^2} \frac{\sqrt{-g}} {g_{tt}g_{xx}}F_{tx}^2. \label{tildeFzy}
\end{eqnarray}
If we add the magnetic field to the p-wave superconductor
\begin{equation}
A(z)= A_t^3(z)\tau^3 dt + (A_x^1(z)\tau^1+A_x^3(y,z)\tau^3)dx,
\label{p-wave}
\end{equation}
one more duality relation can be written
\begin{equation}
\tilde{F}_{zt}^3 = -\frac{1}{g^2} \frac{\sqrt{-g}} {g_{xx}g_{yy}}F_{xy}^3, \label{tildeFzt}
\end{equation}
that makes the system of eqs.(\ref{tildeFxy}-\ref{tildeFzy},\ref{tildeFzt}) closed and symmetric in terms of the gauge field components on the two sides of the duality.
Here the tilde field strength on the B-side eqs.(\ref{tildeFxy}-\ref{tildeFzy},\ref{tildeFzt}) 
has complimentary components to the one on the E-side,
$g=\frac{1}{\tilde{g}}$, and the Yang-Mills coupling $g$ 
should not be confused with the determinant of the metric $\sqrt{-g}$.
The metric factor arises 
from using the following convention for Levi Civita tensor $\epsilon^{1234}=\frac{1}{\sqrt{-g}}$, 
$\epsilon_{1234}=-\sqrt{-g}$ with $g\equiv {\rm det}g_{\mu\nu}$,
where lifting/lowering of indexes is done as usual, for example 
$\epsilon^{ijkl}g_{1i}g_{2j}g_{3k}g_{4l}=-\epsilon_{1234}$
and $\epsilon^{1234}\epsilon_{1234}=-4!=-24$; therefore for example 
$\epsilon_{xyzt}\sim \sqrt{-g}/\sqrt{g_{tt}g_{zz}}$.   

In order to find the field content in the S-dual frame, we need to use explicit solutions of Yang-Mills equations
in the electric frame. We don't know these solutions in the condensed phase. However, we can look at two asymptotic behavior
of the fields: $AdS_4$ expansion in the UV and $AdS_2$ expansion in the IR throat of the RN-AdS geometry.
As a result we obtain the following field components on the magnetic side
\begin{equation}
\tilde{A}(x,z)= \tilde{A}_t^3(z)\tau^3dt + (\tilde{A}_y^2(z)\tau^2+\tilde{A}_y^3(x,z)\tau^3)dy,
\label{p-wave_dual}
\end{equation}
that we explain further.
Therefore the field strengths are given by 
\begin{eqnarray}
&& \tilde{F}_{xy}^3 = \partial_x \tilde{A}_y^{3},\label{Fxy}\\
&& \tilde{F}_{ty}^1 = - \tilde{A}_t^3\tilde{A}_y^2,\label{Fty}\\
&& \tilde{F}_{zy}^2 = \partial_z \tilde{A}_y^2,\label{Fzy}
\end{eqnarray}
that we use in the duality relations eqs.(\ref{tildeFxy}-\ref{tildeFzy}). 
Performing the necessary substitutions,
the duality relations can be written as
\begin{eqnarray}
&& \partial_x\tilde{A}_y^3 = \frac{1}{g^2} \frac{\sqrt{-g}} {g_{tt}g_{zz}} \partial_z A_t^3, \label{duality111}\\
&& \partial_zA_x^1 =  \frac{1}{\tilde{g}^2} \frac{g_{zz}g_{xx}} {\sqrt{-g}}\tilde{A}_t^3\tilde{A}_y^2, 
\label{duality222}\\
&& \partial_z\tilde{A}_y^2 = -\frac{1}{g^2} \frac{\sqrt{-g}} {g_{tt}g_{xx}}A_t^3A_x^1, \label{duality333}
\end{eqnarray}
where $g=1/\tilde{g}$.
When the magnetic field is added to the p-wave superconductor eq.(\ref{p-wave}),
additional duality relation is written
\begin{equation}
\partial_y A_x^3 = -\frac{1}{\tilde{g}^2} \frac{g_{xx}g_{yy}}{\sqrt{-g}} \partial_z \tilde{A}_t^3, \label{duality444}
\end{equation}
In what follows we show that equations (\ref{duality111},\ref{duality444}) 
map the magnetic field 
to the charge density on the opposite side of duality, and equations
(\ref{duality222},\ref{duality333}) relate the v.e.v.'s and the sources on the electric and magnetic sides. 
Therefore eqs.(\ref{duality222},\ref{duality333}) are the duality relations 
for the condensates in the electric and magnetic systems.

The gauge fields which satisfy the equations of motion should also satisfy
the duality relations. EOM are invariant under the S-duality transformations. Next we summarize
the equations of motions on electric and magnetic sides of the duality. 
The Yang-Mills equation for the non-abelian SU(2) gauge fields is 
\begin{equation}
\nabla_{\mu}F^{a\mu\nu} = -\epsilon^{abc}A_{\mu}^b F^{c\mu\nu},
\label{Yang-Mills}
\end{equation}
which becomes for the gauge field components $A_t^3(z)$ and $A_x^1(z)$ from eq.(\ref{p-wave}) 
on the electric side 
\begin{eqnarray}
\frac{1}{\sqrt{-g}}\partial_z\left(\frac{\sqrt{-g}}{g_{zz}g_{tt}}F_{zt}^{3}\right) &=& -\frac{1}{g_{tt}g_{xx}}A_x^1F_{xt}^{2},\\
\frac{1}{\sqrt{-g}}\partial_z\left(\frac{\sqrt{-g}}{g_{zz}g_{xx}}F_{zx}^{1}\right) &=& \frac{1}{g_{tt}g_{xx}}A_t^3 F_{tx}^{2},
\label{Yang-Mills_E}
\end{eqnarray}
respectively, with $g\equiv {\rm det}g$ and the metric notation is given by eq.(\ref{metric}). 
Equations of motion on the electric side, eqs.(\ref{Yang-Mills_E}), are written explicitly 
\begin{eqnarray}
\partial_z^2 A_t^{3}+\frac{g_{zz}g_{tt}}{\sqrt{-g}}\partial_z\left(\frac{\sqrt{-g}}{g_{zz}g_{tt}}\right)\partial_z A_t^{3} 
-\frac{g_{zz}}{g_{xx}}(A_x^1)^2 A_t^3 &=& 0,\label{temp1}\\
\partial_z^2 A_x^{1}+\frac{g_{zz}g_{xx}}{\sqrt{-g}}\partial_z\left(\frac{\sqrt{-g}}{g_{zz}g_{xx}}\right)\partial_z A_x^{1} 
-\frac{g_{zz}}{g_{tt}}(A_t^3)^2 A_x^1 &=& 0.\label{cond1}
\end{eqnarray}
The Yang-Mills equations eq.(\ref{Yang-Mills}) are written for the gauge field components 
$\tilde{A}_y^3(z,x)$, $\tilde{A}_t^3(z)$ and $\tilde{A}_y^2(z)$ eq.(\ref{p-wave_dual}) on the magnetic side 
\begin{eqnarray}
\frac{1}{\sqrt{-g}}\partial_z\left(\frac{\sqrt{-g}}{g_{zz}g_{yy}}\tilde{F}_{zy}^3\right) &=& 0,\\
\frac{1}{\sqrt{-g}}\partial_x\left(\frac{\sqrt{-g}}{g_{xx}g_{yy}}\tilde{F}_{xy}^3\right) &=& 0,\\
\frac{1}{\sqrt{-g}}\partial_z\left(\frac{\sqrt{-g}}{g_{tt}g_{zz}}\tilde{F}_{zt}^3\right) &=&  \frac{1}{g_{tt}g_{yy}}
\tilde{A}_y^2\tilde{F}_{yt}^1,\\
\frac{1}{\sqrt{-g}}\partial_z\left(\frac{\sqrt{-g}}{g_{zz}g_{yy}}\tilde{F}_{zy}^2\right) &=& -\frac{1}{g_{tt}g_{yy}}
\tilde{A}_t^3\tilde{F}_{ty}^1,
\end{eqnarray}
with the metric given by eq.(\ref{metric}).
Equations of motion on the magnetic side become
\begin{eqnarray}
\partial_z^2\tilde{A}_y^3 + \frac{g_{zz}g_{yy}}{\sqrt{-g}}
\partial_z\left(\frac{\sqrt{-g}}{g_{zz}g_{yy}}\right)\partial_z\tilde{A}_y^3 &=& 0,
\\
\partial_x^2\tilde{A}_y^3+\frac{g_{xx}g_{yy}}{\sqrt{-g}}
\partial_x\left(\frac{\sqrt{-g}}{g_{xx}g_{yy}}\right)\partial_x\tilde{A}_y^3 &=& 0,\\
\partial_z^2\tilde{A}_t^3+\frac{g_{tt}g_{zz}}{\sqrt{-g}}
\partial_z\left(\frac{\sqrt{-g}}{g_{tt}g_{zz}}\right)\partial_z\tilde{A}_t^3
-\frac{g_{zz}}{g_{yy}}(\tilde{A}_y^2)^2\tilde{A}_t^3 &=& 0,\label{temp2}
\\
\partial_z^2\tilde{A}_y^2+\frac{g_{zz}g_{yy}}{\sqrt{-g}}
\partial_z\left(\frac{\sqrt{-g}}{g_{zz}g_{yy}}\right)\partial_z\tilde{A}_y^2
-\frac{g_{zz}}{g_{tt}}(\tilde{A}_t^3)^2\tilde{A}_y^2 &=& 0.\label{cond2}
\end{eqnarray}
We are able to find analytic solutions of the equations of motion in the two limiting cases, near the boundary and around the horizon of the black hole at small temperatures. Then we verify that the duality relations are satisfied by these solutions.
In what follows we outline the asymptotic behavior of the AdS-Reissner-Nordstrom black hole metric.
The AdS-RN black hole in $(3+1)$-dimensions is  
\begin{eqnarray}
ds^2 &=& \frac{r^2}{R^2}(-fdt^2+dx^2+dy^2)+\frac{R^2}{r^2}\frac{dr^2}{f},\label{}\\
f &=& 1+\frac{3r_{\star}^4}{r^4}-\frac{r_0^3+3r_{\star}^4/r_0}{r^3},
\label{AdS-RN}
\end{eqnarray}
where the electric charge of the black hole is $q=\sqrt{3}r_{\star}^2$, and $A_t=\frac{\mu}{r}$ with $\mu\sim q$,
and the radius of the black hole horizon is $r_0$, $f(r_0)=0$. 

In the UV, $r\rightarrow \infty$, the red shift factor $f\approx 1$ and the metric 
becomes asymptotically $AdS_4$ 
\begin{equation}
ds^2 = \frac{r^2}{R^2}(-dt^2+dx^2+dy^2)+\frac{R^2}{r^2}dr^2,
\label{}
\end{equation}
with the radius $R$.
Introducing $z=\frac{R^2}{r}$, we have
\begin{equation}
ds^2 = \frac{R^2}{z^2}(-dt^2+dx^2+dy^2+dz^2).
\label{AdS4-metric-asymptotic}
\end{equation}

In the IR, near the black hole horizon $r\rightarrow r_0$, and at low enough temperatures $T/\mu \ll 1$,
the red shift factor has the double zero
\begin{equation}
f\approx 6\frac{(r-r_{\star})^2}{r_{\star}^2}\left(1-\frac{(r_0-r_{\star})^2}{(r-r_{\star})^2}\right),
\end{equation}
where the expansion is carried out in two small parameters $(r-r_{\star})/r_{\star}\ll 1$ and 
$(r_0-r_{\star})/r_{\star}\ll 1$.
Due to the double zero the metric has the $AdS_2$ behavior with a small correction $\sim (r_0-r_{\star})^2$
\begin{equation}
ds^2 = -\frac{(r-r_{\star})^2}{R_2^2}\left(1-\frac{(r_0-r_{\star})^2}{(r-r_{\star})^2}\right)dt^2
+\frac{R_2^2}{(r-r_{\star})^2\left(1-\frac{(r_0-r_{\star})^2}{(r-r_{\star})^2}\right)}dr^2
+\frac{r_{\star}^2}{R^2}(dx^2+dy^2),
\label{}
\end{equation}
with the radius $R_2=\frac{R}{\sqrt{6}}$. In extremal case, $T=0$, the two radii coincide $r_0=r_{\star}$,
and the correction $\sim (r_0-r_{\star})^2$ vanishes. 
Introducing $z=\frac{R_2^2}{r-r_{\star}}$ and $z_0=\frac{R_2^2}{r_0-r_{\star}}$,
we have the leading $AdS_2$ behavior near the horizon at small temperatures $T=\frac{1}{2\pi z_0}\ll 1$
\begin{equation}
ds^2=-\frac{R_2^2}{z^2}\left(1-\frac{z^2}{z_0^2}\right)dt^2+\frac{R_2^2}{z^2\left(1-\frac{z^2}{z_0^2}\right)}dz^2
+\frac{r_{\star}^2}{R^2}(dx^2+dy^2),
\label{AdS2-metric-asymptotic}
\end{equation} 
where the correction $\sim z^2/z_0^2$ to the $AdS_2$ geometry is due to the small but nonzero temperature.   
In what follows we use the two limiting cases with the metric given by 
eqs.(\ref{AdS4-metric-asymptotic},\ref{AdS2-metric-asymptotic}) 
to solve EOM for the gauge fields and to test the duality conditions.

\subsection{UV asymptotics: $AdS_4$}

First we consider the UV limit with pure $AdS_4$ metric eq.(\ref{AdS4-metric-asymptotic}) at small $z\sim 0$
\begin{equation}
ds^2 = \frac{R^2}{z^2}(-dt^2+dz^2+dx^2+dy^2),
\label{metricADS4}
\end{equation}
where $R$ is the AdS radius.

A known analytic solution of Einstein and Yang-Mills equations is 
the AdS Reissner-Nordstr\"om (RN) black hole with the electric charge $q$, where the vector field is $A_t^3=\mu - qz$
and $A_x^1=0$. There is another solution, a hairy RN black hole, which describes a condensed phase and becomes preferable in some parameter range. There, the metric and the behavior of the scalar potential $A_t^3$ are modified mainly in the IR region by the vector potential $A_x^1$ which acquires a radial profile. However an asymptotic form of the solutions near the AdS boundary remains unchanged to the leading order. Yang-Mills equations in the $AdS_4$ are
\begin{eqnarray}
\partial_z^2 A_t^{3} + (A_x^1)^2 A_t^3 &=& 0,\label{EOM1}\\
\partial_z^2 A_x^{1} + (A_t^3)^2 A_x^1 &=& 0.\label{EOM2}
\end{eqnarray}
Therefore we can write the following asymptotic behavior near the AdS boundary in the probe limit
\begin{eqnarray}
A_t^3 &=& \mu - q z + O(z^2),\\
A_x^1 &=& a_x^{(0)} + a_x^{(1)}z +O(z^2),
\end{eqnarray}
where, we can read off according to the AdS-CFT dictionary the physical terms of the boundary CFT.
In CFT terms, $\mu$ is the $U(1)_3$ chemical potential, and $q$ is
the electric charge density. From the asymptotic expansion of the vector potential, the AdS-CFT dictionary says that in the condensed phase 
where $U(1)_3$ is spontaneously broken,
the source term is zero $a_x^{(0)}=0$ and the condensate is given by the v.e.v. $a_x^{(1)}\neq 0$, 
and in the normal phase $a_x^{(1)}=0$ and $a_x^{(0)}\neq 0$.

In the normal phase, $a_x^{(1)}=0$, from eqs.(\ref{tildeFxy}-\ref{tildeFzy}) the only non-vanishing component of the field strength 
$\tilde{F}_{\mu\nu}$
in the magnetic frame is 
\begin{equation}
\tilde{F}_{xy}^3=-q=const,
\end{equation} 
where we absorbed the coupling into redefining the solution $A_t^3$.
Therefore the S-dual of the $U(1)_3$ charged black hole is a state with $U(1)_3$ magnetic field 
\begin{equation}
\tilde{A}_y^3 =-qx.
\end{equation}  
This result probably holds for a backreacted solution.

In the condensed phase, $a_x^{(1)}\neq 0$, the dual field strengths are from eqs.(\ref{tildeFxy}-\ref{tildeFzy}) 
\begin{eqnarray}
\tilde{F}_{xy}^3 &=& -q+O(z),\label{SdualF1}\\
\tilde{F}_{ty}^1 &=& \tilde{g}^2a_x^{(1)}+O(z),\label{SdualF2}\\
\tilde{F}_{zy}^2 &=& \tilde{g}^2a_x^{(1)}\mu z+O(z^2).
\label{SdualF3}
\end{eqnarray}   
In order to see which operators are switched on in the S-dual frame, we need to find the gauge field 
$\tilde{A}_{\mu}^a$ corresponding to the field strength given by eq.(\ref{SdualF1}-\ref{SdualF3}). 
We work in the radial gauge
\begin{equation}
\tilde{A}_z^a=0,
\label{gauge}
\end{equation}
with $a=1,2,3$. 
In eq.(\ref{SdualF1}), the field strength $\tilde{F}_{xy}^3$ 
\begin{equation}
\tilde{F}_{xy}^3=\partial_x\tilde{A}_y^3=-q
\tilde{A}_y^3=-qx,
\end{equation}
is easily integrated to give
\begin{equation}
\tilde{A}_y^3=-qx,
\end{equation}
which yields the magnetic field perpendicular to the $(x,y)$-plane with CFT. In order to integrate $\tilde{F}_{ty}^1$
we assume a stationary condition: $\partial_t(...)=0$, and no breaking of homogeneity in $(x,y)$ directions: no $1/x$, $1/y$ terms
in the potential $\tilde{A}_{\mu}^a$. This yields
\begin{equation}
\tilde{F}_{ty}^1=
\partial_t\tilde{A}_y^1-\partial_y\tilde{A}_t^1+(\tilde{A}_t^2\tilde{A}_y^3-\tilde{A}_t^3\tilde{A}_y^2)=
\tilde{g}^2a_x^{(1)}=const.
\end{equation}
Here the first term is forbidden by stationarity; the second term is zero, otherwise $\tilde{A}_t^1\sim y$ and 
the field strength $\tilde{F}_{ty}^2\sim -\tilde{A}_t^1\tilde{A}_y^3\sim xy$ will be induced;
solving for the third term, it would need $\tilde{A}_t^2\sim -\tilde{g}^2a_x^{(1)}/qx$
and hence it will break homogeneity, $\tilde{F}_{tx}^2\sim 1/x^2$; therefore only the fourth term is left,    
\begin{equation}
\tilde{A}_t^3\tilde{A}_y^2=-\tilde{g}^2a_x^{(1)}.
\label{condensate}
\end{equation}
Eq.(\ref{condensate}) can be solved if a new chemical potential $\tilde{\mu}$ is introduced,
\begin{eqnarray}
\tilde{A}_t^3 &=& \tilde{\mu}+O(z),\\
\tilde{A}_y^2 &=& -\frac{\tilde{g}^2a_x^{(1)}}{\tilde{\mu}}.
\end{eqnarray}
We see that $\tilde{A}_y^2$ explicitly breaks $U(1)_3$ and $SO(2)$ spacial rotations.
Integrating the last field strength in eq.(\ref{SdualF3}) gives 
\begin{equation}
\tilde{F}_{zy}^2=
\partial_z\tilde{A}_y^2-\partial_y\tilde{A}_z^2-(\tilde{A}_z^1\tilde{A}_y^3-\tilde{A}_z^3\tilde{A}_y^1)=
\tilde{g}^2a_x^{(1)}\mu z.
\end{equation}
Here only the first term is nonzero;   
three other terms are zero due to the gauge condition $\tilde{A}_z^a=0$ eq.(\ref{gauge}). Therefore the duality
condition is
 \begin{equation}
\partial_z\tilde{A}_y^2=\tilde{g}^2a_x^{(1)}\mu z,
\label{source}
\end{equation}
that means the small $z$ expansion of $\tilde{A}_y^2$ starts with $\sim z^2$ and there is no linear term $\sim z$. 

Summarizing the duality conditions, we have 
\begin{eqnarray}
\partial_x\tilde{A}_y^3 &=& \tilde{g}^2\partial_z A_t^3,\label{dualityI}\\
\partial_z A_x^1 &=& -\frac{1}{\tilde{g}^2}\tilde{A}_t^3\tilde{A}_y^2  ,\label{dualityII}\\
\partial_z\tilde{A}_y^2 &=& \frac{1}{g^2} A_t^3 A_x^1,\label{dualityIII}
\end{eqnarray}
and when the magnetic field is added to the E-side
\begin{equation}
\partial_y A_x^3 = -g^2\partial_z \tilde{A}_t^3.\label{dualityIV}
\end{equation}
Here eqs.(\ref{dualityI},\ref{dualityIV}) relate magnetic fields and charge densities, 
and eqs.(\ref{dualityII},\ref{dualityIII}) relate condensates and sources
between the two sides of the duality. To see the latter, we write
an expansion for the gauge fields 
\begin{eqnarray}
A_x^1 &=& a_x^{(0)}+a_x^{(1)}z +...,\\
\tilde{A}_y^2 &=& \tilde{a}_y^{(0)}+\tilde{a}_y^{(1)}z +... .
\label{expansionI}
\end{eqnarray}
From the duality condition eq.(\ref{condensate}), eq.(\ref{dualityII}) we obtain the relation between the condensates, 
which are the v.e.v.'s,
\begin{eqnarray}
\tilde{a}_y^{(0)}= - \frac{\tilde{g}^2a_x^{(1)}}{\tilde{\mu}},
\label{condensateI}
\end{eqnarray}
with $\tilde{g}=1/g$.
From the duality condition eq.(\ref{source}), eq.(\ref{dualityIII}) we obtain the relation for the sources
\begin{eqnarray}
a_x^{(0)}=\tilde{a}_y^{(1)}=0,
\label{condensateII}
\end{eqnarray}  
which are switched off.
The duality relations eqs.(\ref{condensateI},\ref{condensateII}) confirms that the source and the v.e.v. 
are interchanged by the S-duality transformation. In the expansion eq.(\ref{expansionI}), 
the source is the leading and the v.e.v. is subleading terms on the electric side, 
while the source is subleading and the v.e.v. is the leading terms on the magnetic side.
Note, that the duality relations for the condensates and v.e.v.'s, as well as for the magnetic fields generated by the charge 
densities,
can be obtained from one another by replacing tilde variables to non-tilde ones and vice versa. They are direct
and inverse S-duality transformations. The minus sign reflects that the matrix $S^2=-1$; 
when the S-duality is applied twice it gives minus sign.  
 
We summarize the components of the vector potential for the p-wave superconductor and its S-dual
in the asymptotic $AdS_4$ space (small $z$) in the Table \ref{Amu}. Left panel represents the standard p-wave superconductor,
right panel is for the case when magnetic field is added.  

\begin{table}[h]
\begin{center}
\begin{tabular}{|c|c|}
\hline
{\rm E\;side} & {\rm B\;side}\\
\hline\hline
       & $\tilde{A}_y^3 = -qx$ \\
\hline
 $A_t^3 = \mu-qz$  & $\tilde{A}_t^3=\tilde{\mu}-\tilde{q}z$\\ 
 $A_x^1 = a_x^{(1)}z$ & $\tilde{A}_y^2=-\frac{\tilde{g}^2a_x^{(1)}}{\tilde{\mu}}$ \\
   \hline   
\end{tabular}
\hspace{2cm}
\begin{tabular}{|c|c|}
\hline
{\rm E\;side} & {\rm B\;side}\\
\hline\hline
$A_x^3 = \tilde{q}y$       & $\tilde{A}_y^3 = -qx$ \\
\hline
 $A_t^3 = \mu-qz$  & $\tilde{A}_t^3=\tilde{\mu}-\tilde{q}z$\\ 
 $A_x^1 = a_x^{(1)}z$ & $\tilde{A}_y^2=-\frac{\tilde{g}^2a_x^{(1)}}{\tilde{\mu}}$ \\
   \hline   
\end{tabular}
\caption{Components of the gauge field for the p-wave superconductor and its S-dual in the $AdS_4$.
Left table represents the standard holographic p-wave superconductor, right panel is when magnetic field is added.}
\label{Amu}
\end{center}
\end{table}

In the Table \ref{Amu}, the $U(1)_3$ charge density in the p-wave superconductor maps 
to the magnetic field in the S-dual frame (left panel), and vice versa the magnetic field in the p-wave superconductor
setting maps to the charge density in the dual frame (right panel). 

As discussed in refs.(\cite{Witten},\cite{Rangamani}), performing the S-dual transformation on the abelian gauge theory
in the $AdS_4$ bulk corresponds to imposing two boundary conditions: Dirichlet (standard) on electric side and Neumann (modified) on magnetic side. There is some additional effort in extrapolating the results for the abelian theory, where
the Dirichlet and Neumann boundary conditions are simply exchanged under S-duality, 
to the non-abelian case (\cite{Rangamani}). However,
in the p-wave superconductor and in its S-dual the $SU(2)$ symmetry is broken  down to $U(1)_3$ by the chemical potential in the third color direction, and formation of condensates happens in the abelian subgroup. The two boundary conditions are characterized in terms of the fall-off conditions of the gauge field near the boundary: when the leading/subleading term is fixed while the subleading/leading term is allowed to fluctuate gives Dirichlet/Neumann boundary conditions (\cite{Rangamani}).
As is familiar from AdS/CFT, in the standard quantization (Dirichlet b.c.) the leading behavior acts as a source
for the conserved current operator and the subleading behavior gives v.e.v. provided source is switched off.
In the modified (alternative) quantization (Neumann b.c.) the roles of source and v.e.v. are interchanged \cite{McGreevy}.
Therefore, as  the expectation value $\langle a_x^{(1)} \rangle$ is the superconducting condensate, in the S-dual frame 
we associate $\langle \tilde{g}^2a_x^{(1)}/\tilde{\mu}\rangle$ with the magnetic condensate, Table \ref{Amu}.

We can speculate about the physical meaning of a new chemical potential $\tilde{\mu}$ in the S-dual frame. 
Depending on the context, it may reflect the density of magnetic monopoles. 
On the electric side,
the conserved quantity of the boundary theory (conserved current $J^{\mu}$) corresponds to electric charge in the bulk. On the S-dual side, the net magnetic charge corresponds to a conserved quantity in the boundary theory (\cite{Witten})
(while the associated with $a_x^{(0)}$ current vanishes at every point $\langle J^{\mu}\rangle =0$). The difference also arises
in CFT's that there are states charged under the global gauge group on the E-side and therefore the Goldstone modes are produces as this symmetry is broken, while on B-side there is a Gauss' law instead and no Goldstone modes arise.

There is the following pattern of breaking the non-abelian gauge group and the spacial $(x,y)$ rotational symmetry in the p-wave superconductor
\begin{eqnarray}
&& SU(2)\xrightarrow{A_t^3} U(1)_3 \xrightarrow{A_x^1} {\rm nothing},\\
&& SO(2)\xrightarrow{A_t^3} SO(2) \xrightarrow{A_x^1} {\rm nothing},
\end{eqnarray}
where the chemical potential $\mu$ 
which is introduced by the the boundary value of the component $A_t^3$,
breaks the $SU(2)$ symmetry down to the diagonal subgroup $U(1)$ which is generated by $\tau^3$.
In order to study the transition to the superconducting state, we allow solutions with non-zero $\langle J_x^1\rangle $ and therefore the nonzero dual gauge field 
$A_x^1$. This solution breaks not only color $U(1)$, but also the spacial rotations $SO(2)$.
The symmetry breaking pattern for the S-dual of p-wave superconductor is
\begin{eqnarray}
&& SU(2)\xrightarrow{\tilde{A}_y^3,\tilde{A}_t^3} U(1)_3 \xrightarrow{\tilde{A}_y^2} {\rm nothing\; (up\; to\; discrete)},\\
&& SO(2)\xrightarrow{\tilde{A}_y^3,\tilde{A}_t^3} SO(2) \xrightarrow{\tilde{A}_y^2} {\rm nothing\; (up\; to\; discrete)},
\end{eqnarray}
where the magnetic field present through the gauge field $\tilde{A}_y^3\sim x$ breaks the non-abelian
gauge group $SU(2)$ down to the diagonal subgroup $U(1)$, but it does not break $(2+1)$-dimensional rotations $SO(2)$. 
The superconducting phase with $A_1^x$ maps to the S-dual state with a nonzero v.e.v. of an operator with gravity dual
$\tilde{A}_y^2$, while normal phases describe CFT's at nonzero density and nonzero magnetic field.    

There is a spontaneous breaking of $U(1)_3$ symmetry by the corresponding v.e.v.
in both p-wave SC and its S-dual.

We can consider properties of the condensate with respect to discrete symmetries. In the parity transformation,
either one or three coordinates change the sign. We adopt the former,
\begin{equation}
P: x\rightarrow -x,\; y\rightarrow y,\; z\rightarrow z,\; t\rightarrow t.
\label{parity_transform}
\end{equation}
We summarize properties of the p-wave superconductor and its S-dual with respect to the parity transformation
in the Table \ref{Parity}.

\begin{table}[h]
\begin{center}
\begin{tabular}{|ccc|ccc|}
\hline
&{\rm E\;side} &&&{\rm B\;side}&\\
\hline\hline
$\mu$ & $q$ & $a_x^{(1)}$ & $\tilde{\mu}$ & $\tilde{q}$ & $\frac{\tilde{g}^2a_x^{(1)}}{\tilde{\mu}}$\\
\hline
$+$ & $+$ & $-$ & $+$ & $+$ & $+$\\
\hline
\end{tabular}
\caption{Properties of the p-wave superconductor and its S-dual under the parity transformation.} 
\label{Parity}
\end{center}
\end{table}
             
From eq.(\ref{parity_transform}), the components of the vector potential $A_t^3,\tilde{A}_t^3,\tilde{A}_y^2,\tilde{A}_y^3$
are P-even and only $A_x^1$ is P-odd. Also, $\epsilon$-symbol in the S-duality transformation changes the parity, that should be taken into account for the components $\tilde{A}_y^2,\tilde{A}_y^3$ calculated as S-dual, as opposed to introducing $\tilde{A}_t^3$. Indeed 
the magnetic field in the S-dual frame is $B\sim \epsilon q$ which is P-odd, and C-odd. 

On the E-side, the p-wave superconducting condensate is vector, while on the B-side the condensate is pseudovector.

Next we check that the gauge fields given in Table \ref{Amu}, which are solutions of equations of motion, 
satisfy the duality relations eqs.(\ref{dualityI}-\ref{dualityIII}).
Equations of motion on the electric side are given in (\ref{EOM1},\ref{EOM2}).
Equations of motion on the magnetic side are 
\begin{eqnarray}
\partial_z^2\tilde{A}_y^3= \partial_x^2\tilde{A}_y^3 &=& 0,\label{EOM3}\\
\partial_z^2\tilde{A}_t^3
+(\tilde{A}_y^2)^2\tilde{A}_t^3 &=& 0,
\label{EOM4}\\
\partial_z^2\tilde{A}_y^2
+(\tilde{A}_t^3)^2\tilde{A}_y^2 &=& 0.
\label{EOM5}
\end{eqnarray}
To the leading order in small $z$, the gauge field components $A_t^3$, $\tilde{A}_t^3$ (chemical potentials) 
and $\tilde{A_y^3}$ (magnetic field)
satisfy EOM and the first duality condition eq.(\ref{dualityI}). For the gauge field components 
$A_x^1$, $\tilde{A}_y^2$ (condensates) we have the following EOM in the probe limit
\begin{eqnarray}
A^{\prime\prime}+\mu^2 A &=& 0,\\
\tilde{A}^{\prime\prime}+\tilde{\mu}^2 \tilde{A} &=& 0,
\end{eqnarray}
and the duality conditions eqs.(\ref{dualityII},\ref{dualityIII})
\begin{eqnarray}
\frac{1}{g^2}A^{\prime} &=& \tilde{\mu}\tilde{A},\label{duality110}\\
\frac{1}{\tilde{g}^2}\tilde{A}^{\prime} &=& -\mu A,\label{duality220}
\end{eqnarray}    
where we omitted the spacial and the $SU(2)$ gauge group indices, $A_x^1\equiv A$ and $\tilde{A}_y^2\equiv \tilde{A}$,
and $\partial_z A\equiv A^{\prime}$.
Solutions of EOM are
\begin{equation}
A \sim \sin \mu z,\,\, \tilde{A} \sim \cos \tilde{\mu} z, 
\end{equation}
because $A$ and $\tilde{A}$ satisfy the Dirichlet and Neumann boundary conditions, respectively,
and the sources are switched off, 
$A(0)=0$ and $\tilde{A}^{\prime}(0)=0$. Indeed the solutions $\sin(\mu z)$ and $\cos(\tilde{\mu} z)$ 
with appropriate choice of constants of integration satisfy the duality conditions eqs.(\ref{duality110},\ref{duality220}). 
These constants of integration define the condensates on electric and magnetic sides.

\subsection{IR asymptotics: $AdS_2$}

Next we consider the IR limit with $AdS_2\times R^2$ metric eq.(\ref{AdS2-metric-asymptotic}) 
at large $z\rightarrow \infty$
\begin{equation}
ds^2=\frac{R_2^2}{z^2}\left[-\left(1-\frac{z^2}{z_0^2}\right)dt^2+\frac{dz^2}{1-\frac{z^2}{z_0^2}}\right]
+\frac{r_{\star}^2}{R^2}(dx^2+dy^2),
\label{}
\end{equation} 
where $z$ is large. Again we check that the gauge field solutions of EOM satisfy the duality constraints. 
Equations of motion on the electric side are
\begin{eqnarray}
\partial_z^2 A_t^3 +\frac{2}{z}\partial_zA_t^3 -\frac{R^4}{6r_{\star}^2}
\frac{(A_x^1)^2 A_t^3}{z^2\left(1-\frac{z^2}{z_0^2}\right)} &=& 0,\label{}\\
\partial_z^2 A_x^1-\frac{2z}{z_0^2\left(1-\frac{z^2}{z_0^2}\right)}\partial_z A_x^1
+\frac{(A_t^3)^2A_x^1}{\left(1-\frac{z^2}{z_0^2}\right)^2} &=& 0, 
\end{eqnarray}
and EOM on the magnetic side are
\begin{eqnarray}
\partial_z^2 \tilde{A}_y^3 - \frac{2z}{z_0^2\left(1-\frac{z^2}{z_0^2}\right)}\partial_z\tilde{A}_y^3 &=& 0,\label{}\\
\partial_x^2 \tilde{A}_y^3 &=& 0,\label{}\\
\partial_z^2 \tilde{A}_t^3 +\frac{2}{z}\partial_z\tilde{A}_t^3 -\frac{R^4}{6r_{\star}^2}
\frac{(\tilde{A}_y^2)^2 \tilde{A}_t^3}{z^2\left(1-\frac{z^2}{z_0^2}\right)} &=& 0,\label{}\\
\partial_z^2 \tilde{A}_y^2-\frac{2z}{z_0^2\left(1-\frac{z^2}{z_0^2}\right)}\partial_z \tilde{A}_y^2
+\frac{(\tilde{A}_t^3)^2\tilde{A}_y^2}{\left(1-\frac{z^2}{z_0^2}\right)^2} &=& 0,
\end{eqnarray}
where $z/z_0\ll 1$ is a small correction due to a small but nonzero temperature $T=\frac{1}{2\pi z_0}$.
The EOM for the condensate components $A_x^1$ and $\tilde{A}_y^2$ are the same. We look for the two solutions
of EOM which satisfy different boundary conditions.
The duality relations that connect the electric and magnetic sides are  
\begin{eqnarray}
\partial_x\tilde{A}_y^3 &=& -\frac{1}{g^2}\frac{6r_{\star}^2}{R^4}z^2\partial_z A_t^3,\label{dualityI0}\\
\partial_z A_x^1 &=& -\frac{1}{\tilde{g}^2\left(1-\frac{z^2}{z_0^2}\right)} \tilde{A}_t^3\tilde{A}_y^2,\label{dualityII0}\\
\partial_z \tilde{A}_y^2 &=& -\frac{1}{g^2\left(1-\frac{z^2}{z_0^2}\right)} A_t^3 A_x^1.\label{dualityIII0}
\end{eqnarray}
To the leading order in $O(1/z)$, solutions for the temporal components in the probe limit are
\begin{eqnarray}
A_t^3 = \frac{\mu}{6z}\left(1-\frac{z}{z_0}\right),\,
\tilde{A}_t^3 = \frac{\tilde{\mu}}{6z}\left(1-\frac{z}{z_0}\right),\label{temporal}
\end{eqnarray}
with $\partial_zA_t^3 = -\frac{\mu}{6z^2}$. It produces the duality relation eq.(\ref{dualityI0})
\begin{equation}
\partial_x\tilde{A}_y^3=\frac{r_{\star}^2}{g^2R^4}\mu,
\end{equation}
that gives the magnetic field perpendicular to the $(x,y)$ plane
\begin{equation}
\tilde{A}_y^3 = \frac{r_{\star}^2}{g^2R^4}\mu x.
\end{equation}
Using solutions for the temporal components eq.(\ref{temporal}), EOM for the condensate components are
\begin{eqnarray}
A^{\prime\prime} - \frac{2zA^{\prime}}{z_0^2\left(1-\frac{z^2}{z_0^2}\right)}
+\left(\frac{\mu}{6}\right)^2\frac{A}{z^2\left(1+\frac{z}{z_0}\right)^2} &=& 0,\label{}\\
\tilde{A}^{\prime\prime} - \frac{2z\tilde{A}^{\prime}}{z_0^2\left(1-\frac{z^2}{z_0^2}\right)}
+\left(\frac{\tilde{\mu}}{6}\right)^2\frac{\tilde{A}}{z^2\left(1+\frac{z}{z_0}\right)^2} &=& 0,\label{}
\end{eqnarray}
and the duality relations eqs.(\ref{dualityII0},\ref{dualityIII0}) are
\begin{eqnarray}
A^{\prime} &=& \frac{1}{\tilde{g}^2}\frac{\tilde{\mu}}{6z\left(1+\frac{z}{z_0}\right)}\tilde{A},\label{dualityII1}\\
\tilde{A}^{\prime} &=& \frac{1}{g^2}\frac{\mu}{6z\left(1+\frac{z}{z_0}\right)}A,\label{dualityIII1}
\end{eqnarray}
where we omitted the space and group indices, $A_x^1\equiv A$ and $\tilde{A}_y^2\equiv \tilde{A}$, 
and $\partial_z A\equiv A^{\prime}$. Solutions of EOM in the leading order of $O(1/z)$ and $O(z/z_0)$ are
\begin{eqnarray}
&& A\sim \frac{1}{z}\left(1+\frac{z}{z_0}\right)+O\left(\frac{1}{z^2}\right),\\
&& \tilde{A}\sim \frac{1}{z}\left(1+\frac{z}{z_0}\right)+O\left(\frac{1}{z^2}\right).
\end{eqnarray}
Because $A^{\prime}\sim -\frac{1}{z^2}$, these solutions satisfy 
the duality relations eq.(\ref{dualityII1},\ref{dualityIII1}) in each order of perturbation theory in $1/z$.

Thus we showed analytically that in the UV and IR limits, solutions of EOM on electric and magnetic sides are related
by the duality conditions. In the next section we solve EOM and check the duality constraint 
in the holographic bulk numerically.

\section{Numerical solutions and duality mapping between them} \label{sec:3}

In section \ref{sec:2} we demonstrated the $SL(2,Z)$ invariance on the level of the non-Abelian $SU(2)$ action. Also we demonstrated the S-duality
for the equations of motion in the asymptotic UV and IR regimes. In general the S-duality cannot be traced at the EOM level due to the covariant derivative that introduces the gauge field instead of the field strength for which the duality relation is written. Therefore, we solve the EOM directly and show that the physical solutions on different sides are connected by the S-duality relation.

We look for the nontrivial solutions of EOM describing the condensates: 
$A_{x}^1$ in eq.(\ref{cond1}) on the electric side and $\tilde{A}_{y}^2$ in eq.(\ref{cond2}) on the magnetic side. 
The gauge fields $A_{x}^1$ and $\tilde{A}_{y}^2$ satisfy the same equations. Therefore we will be looking for two nontrivial condensate solutions of one EOM. As shown in section \ref{sec:2}, one imposes for the solutions
on E- and B-sides two different UV boundary conditions, Dirichlet and Neumann b.c. respectively. 
This situation is known to arise in holographic superconductor
that is built using the bulk scalar field (the s-wave holographic superconductivity) \cite{NewInstability1,NewInstability2,NewInstability3,NewInstability4}
and in the Sakai-Sugimoto model \cite{NewInstability0}. In the former case, one obtains a "standard" hairy solution at a threshold charge density, i.e. for $\mu\geq \mu_c$, using the Dirichlet b.c. (standard quantization). Also a "new" instability occurs at small charge density --  scalar hair, with Neumann b.c.
(alternative quantization) \cite{NewInstability1,NewInstability2}. In the literature different explanations are given to what
causes a "new" instability \cite{NewInstability1,NewInstability2,NewInstability3,NewInstability4}.
Here we will find two types of instabilities in the holographic p-wave superconductor. 
However, contrary to $AdS_5$ where analytic solution for the "standard" p-wave superconductor
is known \cite{AdS5}, there is no analytic solution in $AdS_4$ and we solve it numerically. 

We use the metric 
\begin{equation}
ds^2 = \frac{1}{z^2}\left(-fdt^2 +\frac{dz^2}{f} +dx_1^2+dx_2^2\right),
\label{general-metric}
\end{equation}  
where the redshift factor for the $AdS_4$-Reissner-Nordstrom black hole is 
\begin{equation}
f = 1 + q^2z^4 -(1+q^2)z^3 =
(1-z)(1+z+z^2-z^3q^2),
\label{AdS-RN1}
\end{equation}
where $q$ is the charge of the black hole, $q=\sqrt{3}r_{\star}^2$. 
Equation (\ref{AdS-RN1}) can be obtained from eq.(\ref{AdS-RN}) 
by rescaling to make $r_0=1$ and $R=1$ and changing the variable $r=1/z$. For the extremal black hole $T=0$,
the redshift factor develops a double zero near the horizon
\begin{equation}
f = 6(1-z)^2,
\end{equation}
and the metric reduces to $AdS_2\times R^2$. As $q=0$, we have the known
metric of the $AdS_4$-Schwarzschild black hole with the redshift factor \cite{Gubser}  
\begin{equation}
f=1-z^3=(1-z)(1+z+z^2)
\label{Sch1}
\end{equation}
In both cases eqs.(\ref{AdS-RN1},\ref{Sch1}), the black hole horizon is at $z=1$, $f(z=1)=0$, and the boundary is at $z=0$.
In this metric eq.(\ref{general-metric}),
the EOM for the magnetic field component $\tilde{A}_y^3$ are
\begin{eqnarray}
\partial_z^2\tilde{A}_y^3 + \frac{f^{\prime}}{f}\partial_z\tilde{A}_y^3 &=& 0,\\
\partial_x^2\tilde{A}_y^3 &=& 0
\end{eqnarray}
and the duality relation takes the form
\begin{equation}
\partial_x\tilde{A}_y^3 = \frac{1}{g^2}\partial_z A_t^3.
\end{equation}
EOM and the duality constraint are solved in the probe limit: $A_t^3=\mu(1-z)$ and the constant magnetic field 
$\tilde{A}_y^3 = - \mu x$. A nontrivial task is to solve the EOM and check the duality for the condensate components.
In the metric given by eq. (\ref{general-metric}), 
the EOM for the temporal $A_t^3$ and condensate $A_x^1$ gauge field components on the electric side are
\begin{eqnarray}
A_t^{\prime\prime}-\frac{(A_x)^2A_t}{f} &=& 0,\label{temporal100} \\
A_x^{\prime\prime}+ \frac{f^{\prime}A_x^{\prime}}{f} + \frac{(A_t)^2A_x}{f^2} &=& 0.\label{condensate100}
\end{eqnarray}
The same system of equations is obtained for the temporal $\tilde{A}_t^3$ 
and the condensate $\tilde{A}_y^2$ components on the magnetic side. 
For now we omit the group indices by the gauge fields, and denote $\partial_z A \equiv A^{\prime}$. 
In the UV at $z=0$, the asymptotic behavior of the solution is
\begin{equation}
A = A^{(0)}+zA^{(1)}+ ...  .
\end{equation} 
To obtain a nontrivial condensate solution we need to switch off the source. The Dirichlet boundary condition implies that the leading
source term is $A^{(0)}$ and the subleading term $A^{(1)}$ is a condensate. 
For the Neumann boundary condition the role of the source and the v.e.v. is interchanged, i.e.
$A^{(1)}$ is the source and $A^{(0)}$ is the v.e.v.. Therefore the UV behavior of the two solutions is
\begin{eqnarray}
&& {\rm Dirichlet\; (E\;side):}\; A^{(0)}=0,\; A^{\prime}(z=0)=A^{(1)}={\rm v.e.v.},\label{}\\
&& {\rm Neumann\;(B\;side):}\; \tilde{A}^{(1)}=0,\; \tilde{A}(z=0)=\tilde{A}^{(0)}={\rm v.e.v.},\label{}
\label{}
\end{eqnarray}  
where we read off the condensates as v.e.v. on both sides of duality.
The duality conditions for the condensate components read
\begin{eqnarray}
A_x^{\prime} &=& \frac{\tilde{A}_t\tilde{A}_y}{\tilde{g}^2f},\label{duality-condition1}\\
\tilde{A}_y^{\prime} &=& -\frac{A_tA_x}{g^2f},\label{duality-condition2}
\end{eqnarray}
that relates the solutions of the EOMs on the electric and magnetic sides  $A_x^1$ and $\tilde{A}_y^2$ with each other. In what follows we consider the probe limit,
where solutions for the temporal gauge components read
\begin{equation}
A_t=\mu(1-z),\,\tilde{A}_t=\tilde{\mu}(1-z).
\end{equation}
Next we check the asymptotic regimes in the UV and IR analytically. 
In the UV, at $z=0$, the redshift factor is $f=1$ for the AdS-RN and the Schwarzschild black holes, thus it is asymptotically an $AdS_4$ metric.
Therefore the asymptotic EOM and the asymptotic duality relations are
\begin{eqnarray}
z \sim 0:
&& A_x^{\prime\prime}+\mu^2A_x = 0,\,\tilde{A}_y^{\prime\prime}+\tilde{\mu}^2\tilde{A}_y = 0,\label{}\\
&& A_x^{\prime} = \frac{\tilde{\mu}}{\tilde{g}^2}\tilde{A}_y,\,\tilde{A}_y^{\prime} = -\frac{\mu}{g^2}A_x,\label{AdualityUV}
\end{eqnarray}
solved by
\begin{equation}
A_x\sim \sin(\mu z),\, \tilde{A}_y\sim \cos(\tilde{\mu}z).
\end{equation}
One can also express, using the duality relation, the dual field $\tilde{A}$ on the B-side through the original one $A$ on the E-side and substitute it in the EOM, 
to make sure that the EOM are satisfied. In the IR, at $z=1$, and at small temperatures the redshift factor
for an AdS-RN black hole is $f=6(1-z)^2$. In the $AdS_2\times R^2$ metric, the EOM and the duality relations are
\begin{eqnarray}
z\sim 1:
&& A_x^{\prime\prime}-\frac{2}{1-z}A_x^{\prime} +\frac{\mu^2}{36(1-z)^2}A_x =0,\\
&& \tilde{A}_y^{\prime\prime}-\frac{2}{1-z}\tilde{A}_y^{\prime} +\frac{\tilde{\mu}^2}{36(1-z)^2}\tilde{A}_y =0,\\
&& A_x^{\prime}=\frac{\tilde{\mu}}{\tilde{g}^2}\frac{\tilde{A}_y}{6(1-z)},\label{duality10}\\
&& \tilde{A}_y^{\prime}=-\frac{\mu}{g^2}\frac{A_x}{6(1-z)}.\label{duality20}
\end{eqnarray}
Because of the second term $A^{\prime}/(1-z)$ in the EOM, an expansion of the solution starts from $(1-z)^2$ to ensure the regularity 
\begin{eqnarray}
A_x &\sim & (1-z)^2+O\left((1-z)^3\right),\\
\tilde{A}_y & \sim & (1-z)^2+O\left((1-z)^3\right),
\end{eqnarray}
that satisfy the duality relations in each order of the expansion. It happens due to the double zero
in the redshift factor $f\sim (1-z)^2$ 
that leads to the duality relation having structure $A^{\prime}\sim \tilde{A}/(1-z)$.
Therefore the S-duality holds in the IR for an AdS-RN black hole for small enough temperatures 
where the metric reduces to an $AdS_2$ throat.
 
To show that the solutions of the EOM satisfy the duality relations in the holographic bulk we resort to a numerical study. 
It is convenient to rewrite the EOM for the condensate component in the form of Riccati equation \cite{Papadimitriou},
that transforms the linear ODE of the second order
into a nonlinear ODE of the first order.\footnote{
In the second order equation $\alpha(x)y^{\prime\prime}+\beta(x)y^{\prime}+\gamma(x)y=0$, 
we make substitution $w=-\frac{y^{\prime}}{\alpha(x)y}$. Then the Riccati
equation is given by 
$w^{\prime}=\alpha(x)w^2+\frac{\alpha^{\prime}(x)-\beta(x)}{\alpha(x)}w+\frac{\gamma(x)}{\alpha^2(x)}$.
}
In this way one needs to specify only one bondary condition instead of two.
Introducing $w=A_x^{\prime}/A_x$ and $\tilde{w}=\tilde{A}_y^{\prime}/\tilde{A}_y$ in eq.(\ref{condensate100}), 
we obtain the following EOM and the duality relation
\begin{eqnarray}
&& w^{\prime}+w^2+\frac{f^{\prime}}{f}w + \frac{A_t^2}{f^2} = 0,\label{w-EOM1} \\
&& \tilde{w}^{\prime}+\tilde{w}^2+\frac{f^{\prime}}{f}\tilde{w} 
+ \frac{\tilde{A}_t^2}{f^2} = 0,\label{w-EOM2} \\
&& w\tilde{w} = - \frac{A_t\tilde{A}_t}{f^2}, \label{w-duality}
\end{eqnarray}
where the metric is given by eq.(\ref{AdS-RN1}) with $f(z)$ specified for the AdS-RN/Schwarzschild BH
and the probe limit solutions
$A_t=\mu(1-z)$ and $\tilde{A}_t=\tilde{\mu}(1-z)$ are used.
The EOM are supplemented by the IR boundary condition 
\begin{eqnarray}
z \sim 1:
&& w(z\sim 1)= w_1(1-z),\label{IRduality100}\\
&& \tilde{w}(z\sim 1)= -\frac{2}{1-z},\label{IRduality200}
\end{eqnarray} 
where $w_1=w^{\prime}(z=1)$ is a constant. 
Note, at $z=1$ the redshift factor is $f=(1-z)(3-q^2)=3(1-z)(1-r_{\star}^4)$ 
for the AdS-RN BH and $f=3(1-z)$ for the Schwarzschild BH.   
This boundary condition
ensures that the condensate gauge field solutions are regular in the IR.  
It corresponds to the following
behavior of the condensate fields in the IR
\begin{eqnarray}
A_x &=& a^{(0)} + a^{(2)}(1-z)^2+...,\\
\tilde{A}_y &=& \tilde{a}^{(2)}(1-z)^2+... .
\end{eqnarray}
There is no boundary condition in the IR apart from the regularity condition. A regular solution is obtained when 
$A^{\prime}(z=1)=0$ in eq.(\ref{condensate100}), that is there is no $(1-z)$ term in the gauge field solution at $z=1$.
To obtain the duality for $w$'s, the original
duality conditions eqs.(\ref{duality-condition1},\ref{duality-condition2})
are $Z_2$ reflected (the no-tilde-variables interchange with the tilde variables) and the relation $\tilde{g}=1/g$ is used.

\begin{figure}[!ht]
\includegraphics[width=0.35\textwidth]{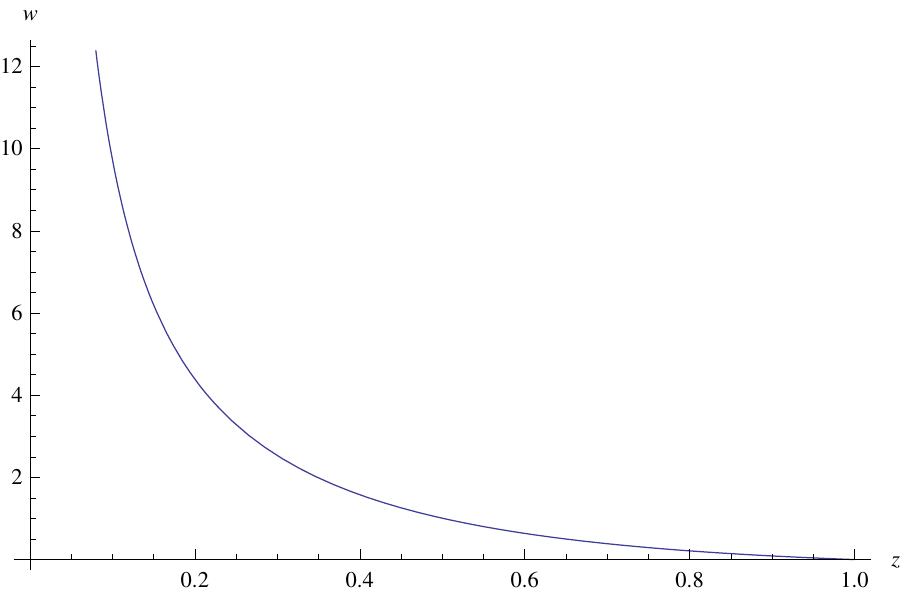}
\includegraphics[width=0.35\textwidth]{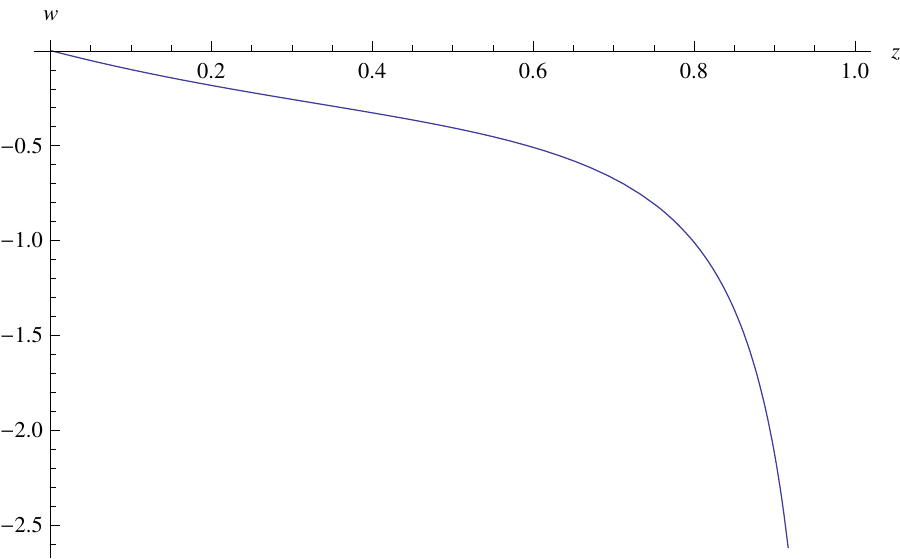}
\caption{Two solutions $w$ of the Riccati equation in the $AdS_4$-Schwarzschild metric: 
a standard solution for $\mu=3.66$, $w_1=0.21$ 
 (left) and a new solution
for $\mu=1.05$, $\tilde{w}_1=-2$ (right).}
\label{Riccati_solution}
\end{figure}

In the IR, the duality relation eq.(\ref{w-duality}) gives 
\begin{equation}
w(z=1)\tilde{w}(z=1) = \frac{\mu\tilde{\mu}}{(3-q^2)^2},\label{IR-duality} 
\end{equation}
for the AdS-RN and Schwarzschild black holes. Due to eqs.(\ref{IRduality100},\ref{IRduality200}), the IR duality
condition eq.(\ref{IR-duality}) gives
\begin{equation}
w^{\prime}(z=1) = -\frac{\mu\tilde{\mu}}{2(3-q^2)^2}.\label{IR-duality100} 
\end{equation}
This equation fixes a constant $w_1=w^{\prime}(z=1)$ in the IR boundary condition eq.(\ref{IRduality100}).

\begin{figure}[!ht]
\includegraphics[width=0.35\textwidth]{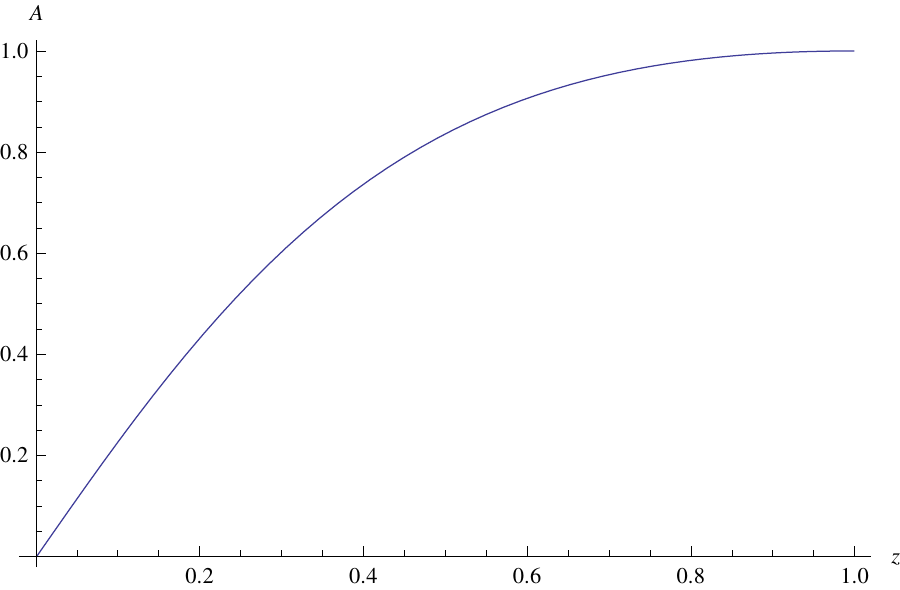}
\includegraphics[width=0.35\textwidth]{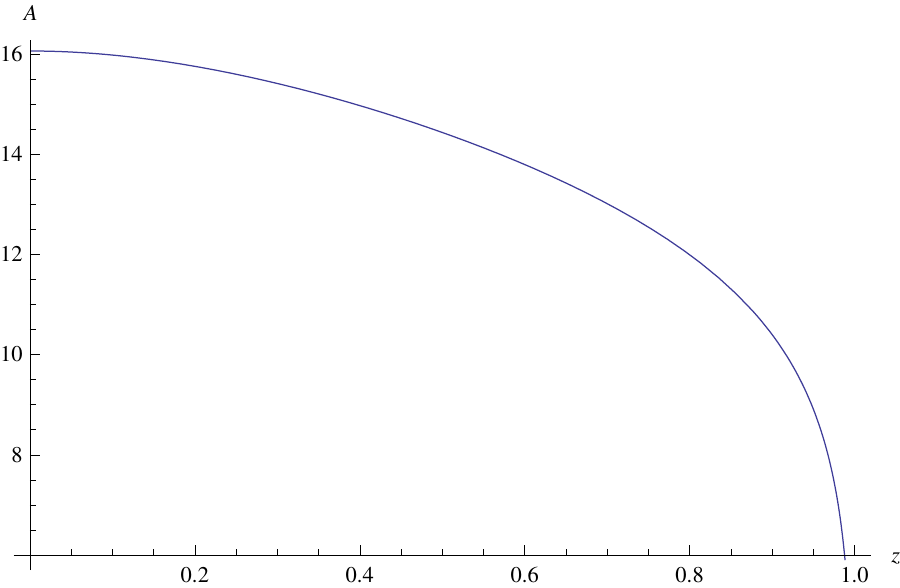}
\caption{Two solutions for the gauge field $A$ of the equation of motion in the $AdS_4$-Schwarzschild metric: 
a standard solution for $\mu=3.656$, $w_1=0.21$  
(left) and a new solution for $\tilde{\mu}=1.05$, $\tilde{w}_1=-2$ (right).}
\label{EOM_solution}
\end{figure}

The UV behavior of the solutions of the Riccati equations is
\begin{eqnarray}
z\sim 0:
&& {\rm Dirichlet\;b.c.\; (E\;side)}\; w(z\sim 0)=\frac{1}{z}\rightarrow \infty,
\label{UV-boundary1}\\
&& {\rm Neumann\; b.c.\; (B\;side)}\; \tilde{w}(z\sim 0)= \tilde{w}_0 z \rightarrow 0,\label{UV-boundary2}
\label{}
\end{eqnarray}
where $\tilde{w}_0=w^{\prime}(z=0)$ is a constant.  
It translates into the following behavior of the gauge fields in the UV
\begin{eqnarray}
A_x &=& A^{(1)}z +...,\label{UVduality300}\\
\tilde{A}_y &=& \tilde{A}^{(0)} + \tilde{A}^{(2)}z^2+...,\label{UVduality400}
\end{eqnarray} 
with the sources being switched off on both sides of the duality.
In the UV, the duality relation eq.(\ref{w-duality}) amounts to
\begin{equation}
w(z=0)\tilde{w}(z=0) = -\mu\tilde{\mu}.\label{UV-duality} 
\end{equation}
Due to eqs.(\ref{UV-boundary1},\ref{UV-boundary2}), the UV duality condition eq.(\ref{UV-duality}) gives 
\begin{equation}
\tilde{w}^{\prime}(z=0) = -\mu\tilde{\mu}.\label{UV-duality100} 
\end{equation}

\begin{figure}[!ht]
\includegraphics[width=0.35\textwidth]{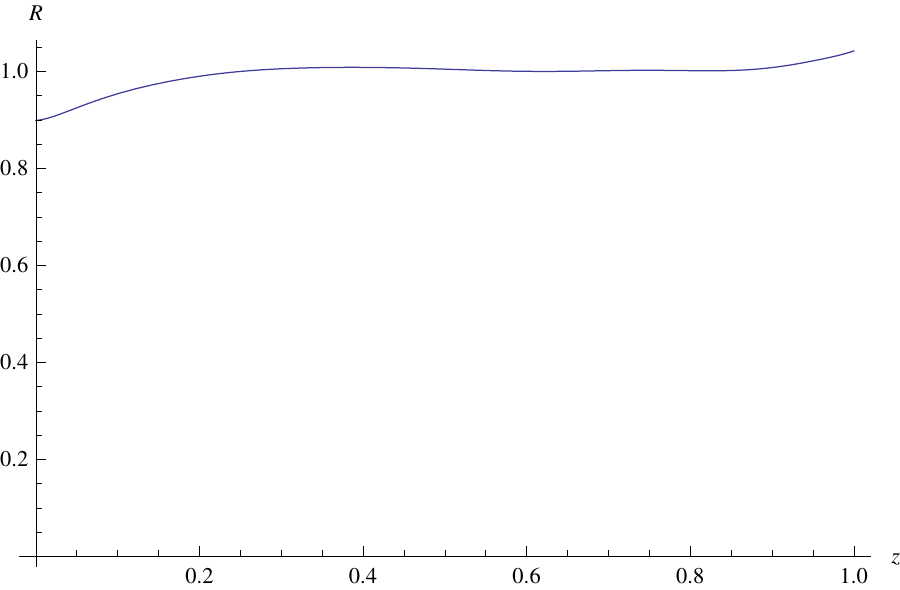}
\caption{Combination $R(z)=-w(z)\tilde{w}(z)(1 + z + z^2-z^3q^2)^2$ for the two solutions
of the Riccati equation in the Schwarzschild metric $q=0$. Combination $R$ being constant means that
the duality relation eq.(\ref{w-duality}) is satisfied in the AdS-RN space.}
\label{Duality_check}
\end{figure}

We rewrite the S-duality equation (\ref{UV-duality}) in the UV using the connection between 
the Riccati variable $w$ calculated at the boundary and the Green's function
\begin{eqnarray}
w(z=0) &=& \left.\frac{A_x^{\prime}}{A_x}\right|_{\rm UV} \sim G_{xx}^{11}(\omega=k=0) \nonumber\\
\tilde{w}(z=0) &=& \left.\frac{\tilde{A}_y^{\prime}}{A_y}\right|_{\rm UV} \sim \tilde{G}_{yy}^{22}(\omega=k=0)
\end{eqnarray}
where the retarded Green's function is
$G_{ij}(\omega)=-i\int d^2xdt{\rm e}^{-i\omega t}\theta(t)\langle [J_i(t),J_j(0)]\rangle$ 
and $J_i$ is the current. The duality relation (\ref{UV-duality}) reads
\begin{equation}
G_{xx}^{11}\tilde{G}_{yy}^{22} = \mu\tilde{\mu}.
\end{equation} 
Therefore, in $2+1$ theory at nonzero density, the $S$-duality transformation 
$E\xrightarrow{S}B$ acts as follows
\begin{equation}
\frac{G(\omega=k=0)}{\mu} \xrightarrow{S} \left[\frac{G(\omega=k=0)}{\mu}\right]^{-1},
\label{Greens}
\end{equation}
where the Green's function is associated with the boundary directions, i.e. $G_{xx}$. Formally identifying
the real part of the retarded Green's function at zero frequency with the superfluid density
$n_s=\operatorname{Re}\left[G^{R}(\omega=k=0)\right]$ \cite{sigmaDC1}-\cite{sigmaDC4}, \cite{Gubser} we can rewrite eq.(\ref{Greens}) 
\begin{equation}
n_s/\mu \xrightarrow{S} \frac{1}{n_s/\mu},
\label{superfluid}
\end{equation}
when the duality transformation is performed. However, the meaning of eq.(\ref{superfluid}) may be obscure
because the superfluid density should be identified with the direction in the isospin space of the conserved charge 
which is $U(1)_3$. We leave eq.(\ref{superfluid}) as a speculative suggestion that can be realized in other models 
at nonzero charge densities where the change of transport coefficients with duality transformation is considered.  

To this end we consider the duality relations for the electrical conductivity which arises 
from the nonabelian current $J_{x,y}^3$ generated by $\tau^3$ component. 
The electrical conductivity is defined through the Ohm's law
\begin{equation}
J_i=\sigma_{ij}E^{j},
\end{equation}
where $E^j$ is an external electric field and $J_i$ is the current generated.
The current $J_{x,y}^3$ is dual to fluctuations
of the $\delta A_{x,y}^3$ fields. Because of the non-abelian Yang-Mills action, the fluctuations in $\delta A_{x,y}^3$
will source other field components. We will keep all the modes which couple at a linearised level.

The gauge field includes the background and fluctuation components $A_i+\delta A_i$. We summarize the background gauge fields
on the two sides of the duality 
\begin{equation}
\left(A_t^3,A_x^1;A_x^3\right)\xrightarrow{S} \left(\tilde{A}_t^3,\tilde{A}_y^2;\tilde{A}_y^3\right).
\label{background}
\end{equation}    
We consider fluctuations that have the same charge as $\delta A_{x,y}^3$ under $U(1)_3$ action. There will be decoupled equations involving the following fluctuation fields on the two sides of the duality 
 \begin{equation}
\left(\delta A_x^3,\delta A_t^2; \delta A_x^1\right) \xrightarrow{S}  
\left(\delta\tilde{A}_y^3,\delta\tilde{A}_t^1;\delta\tilde{A}_y^2\right).
\label{fluctuation}
\end{equation} 
In eqs.(\ref{background},\ref{fluctuation}) semicolon separates the gauge fields responsible for the magnetic fields.   
Additionally the fluctuations $\delta A_z^{1,2}$ arise in coupled equations of motion.
We use a background field gauge transformation to set $\delta A_z^{1,2}=0$ \cite{RobertsHartnoll}.
All fluctuation fields are taken to have an overall time dependence of ${\rm e}^{-i\omega t}$.

Integrating the fields to the UV, we can read off the dual currents and external electric fields. The current
and charge densities are obtained from 
\begin{equation}
F_{z\mu}^a = \langle J_{\mu}^a\rangle + ...,
\end{equation}
where $\mu$ runs over the boundary directions $t,x,y$. The background equilibrium values are 
$\langle J_x^1\rangle=J$, $\langle \tilde{J}_y^2\rangle= \tilde{J}$ and 
$\langle J_t^3\rangle=\rho$, $\langle \tilde{J}_t^3\rangle=\tilde{\rho}$.
The external electric fields are obtained from
\begin{equation}
F_{ti}^a = -E_i^a + ....
\end{equation}
We summarize the duality relations for the background fields in the UV (i.e. omitting the metric factors
and the Yang-Mills coupling constant)
\begin{eqnarray}
\tilde{F}_{zy}^2 &\sim & F_{tx}^2,\label{background1}\\
\tilde{F}_{ty}^1 &\sim & F_{zx}^1,\label{background2}\\
\tilde{F}_{xy}^3 &\sim & F_{zt}^3,\label{background3}\\
\tilde{F}_{zt}^3 &\sim & F_{xy}^3,\label{background4}
\end{eqnarray}   
where the field strengths are
\begin{eqnarray}
\tilde{F}_{zy}^2=\partial_z\tilde{A}_y^2, && F_{tx}^2=A_t^3 A_x^1,\\
\tilde{F}_{ty}^1=-\tilde{A}_t^3\tilde{A}_y^2, && F_{zx}^1=\partial_z A_x^1,\\
\tilde{F}_{xy}^3=\partial_x \tilde{A}_y^3, && F_{zt}^3=\partial_zA_t^3,\\
\tilde{F}_{zt}^3=\partial_z\tilde{A}_t^3, && F_{xy}^3=\partial_yA_x^3,
\end{eqnarray}

The first two equations (\ref{background1},\ref{background2}) provide the relation in the symmetric form for the Green's functions when the duality transformation is done, while in the equations (\ref{background3},\ref{background4}) the charge density generates the magnetic field on the other side of the duality. 

The duality relations for the gauge field fluctuations are written (again omitting the metric factors and the coupling constant) as follows 
\begin{eqnarray}
\tilde{F}_{ty}^3 &\sim & F_{zx}^3,\label{fluctuation1}\\
\tilde{F}_{zy}^3 &\sim & F_{tx}^3,\label{fluctuation2}\\
\tilde{F}_{xy}^2 &\sim & F_{zt}^2,\label{fluctuation3}\\
\tilde{F}_{zt}^1 &\sim & F_{xy}^1,\label{fluctuation4}
\end{eqnarray}    
where the field strengths on the linearised level are
\begin{eqnarray}
\tilde{F}_{ty}^3=-\tilde{E}_y^3=\partial_t\delta\tilde{A}_y^3+\tilde{A}_y^2\delta\tilde{A}_t^1, 
&& F_{zx}^3= J_x^3=\partial_z\delta A_x^3,\\
\tilde{F}_{zy}^3=\tilde{J}_y^3=\partial_z\delta\tilde{A}_y^3, 
&& F_{tx}^3=-E_x^3=\partial_t \delta A_x^3-A_x^1\delta A_t^2,\\
\tilde{F}_{xy}^2=\partial_x \delta\tilde{A}_y^2, && F_{zt}^2=\partial_z\delta A_t^2,\\
\tilde{F}_{zt}^1=\partial_z\delta\tilde{A}_t^1, && F_{xy}^1=\partial_y \delta A_x^1,
\end{eqnarray}
where we simplified $\langle J_x^3\rangle = J_{x}^3$ and $\langle \tilde{J}_y^3\rangle = \tilde{J}_{y}^3$. 

We are interested in the electrical conductivity of the $U(1)$ subgroup of $SU(2)$
generated by $\tau^3$. Therefore we consider currents $J_{x,y}^3$ that result from external sources 
in the $\tau^3$ direction only. Therefore 
we read off the linearised electric response to a time varying external electric field
$E_i^3$ only. However, if we integrate equations of motion to the boundary, we would not obtain electric
fields $E_{i}^{1,2}$, we would obtain a source $\delta A_t^{1,2}$. 
Therefore a gauge transformation in the bulk should be done
that sets the boundary value of $\delta A_t^{2}$, $\delta \tilde{A}_t^{1}$ to zero \cite{RobertsHartnoll}, 
which results in the new scalar potentials 
and the new field strengths. In what follows the specific form of the field strengths is not important for us.
The first two duality equations (\ref{fluctuation1},\ref{fluctuation2}) give relation between the electric
conductivities
\begin{eqnarray}
\sigma_{xx}^3 &=& \frac{ J_x^3}{E_x^3} = \lim_{z\to 0}\frac{F_{zx}^3}{F_{tx}^3},\\
\tilde{\sigma}_{yy}^3 &=& \frac{\tilde{J}_y^3}{\tilde{E}_y^3} = 
\lim_{z\to 0}\frac{\tilde{F}_{zy}^3}{\tilde{F}_{ty}^3},
\end{eqnarray}       
when written in the symmetric form
\begin{equation}
\sigma_{xx}^3\tilde{\sigma}_{yy}^3 =1.
\end{equation}
This relation holds when all the metric factors and coupling constant are restored. Alternatively, 
this equation shows how the electric conductivity transforms when the $S$-duality transformation 
$E\xrightarrow{S}B$ is done
\begin{equation}
\sigma \xrightarrow{S} \frac{1}{\sigma}.
\label{conduct}
\end{equation}
The relation for conductivities was first established for the self-dual CFTs
in \cite{Conductivity} and then in \cite{Conductivity2} for the CFTs where the EM self-duality is lost. 
Later it was shown by W. Witczak-Krempa and S. Sachdev
in \cite{Conductivity2}, that
the particle-vortex or S-duality interchanges the locations of the conductivity zeros and poles.
Specifically, the poles of the dual conductivity
$\tilde{\sigma}\sim 1/\sigma$,
correspond to the zeros of the conductivity $\sigma$ in $\omega$ complex plane and vice versa. It was also shown 
that the S-duality
transformation corresponds to the metal-insulator transition. We consider the equation (\ref{conduct})
as a generalization of the duality relation for conductivities obtained by W. Witczak-Krempa and S. Sachdev 
to theories at nonzero densities. 

The last two duality equations (\ref{fluctuation3},\ref{fluctuation4}) relate the charge density and the magnetic field on the two sides of the duality.

\begin{figure}[!ht]
\includegraphics[width=0.35\textwidth]{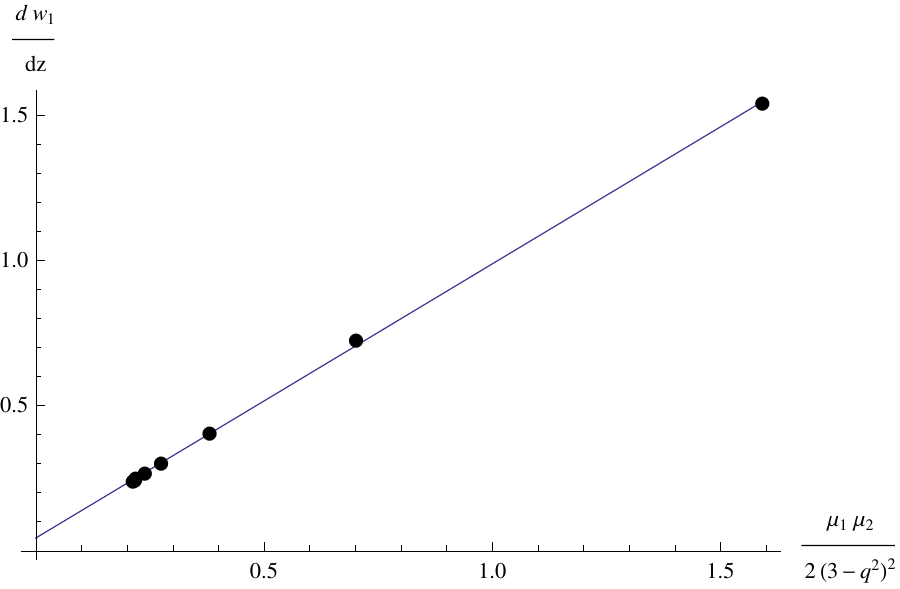}
\includegraphics[width=0.35\textwidth]{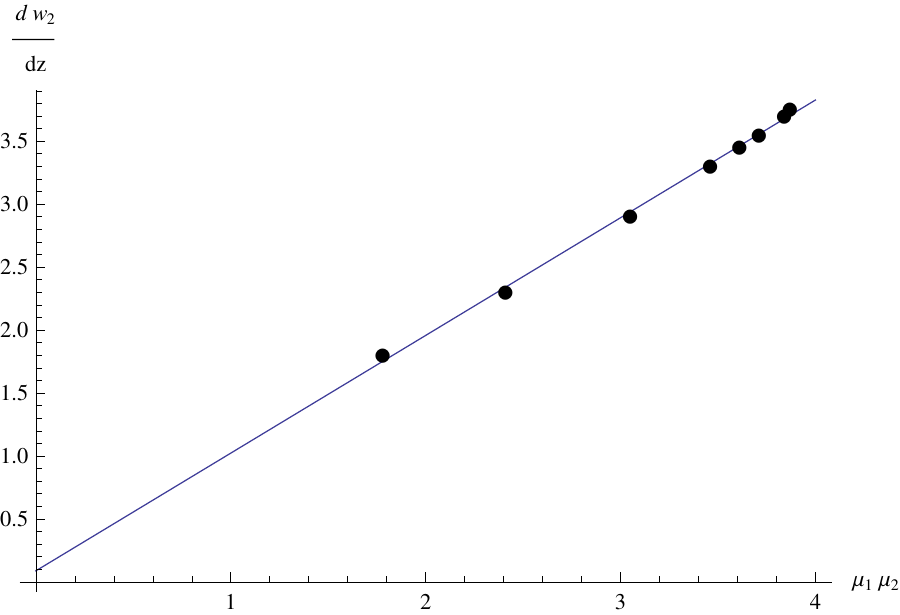}
\caption{The duality relation for the RN-AdS BH metric in the IR and the UV asymptotics
 with the charge increased from $q=0$ to $q=1.5$.
The duality in the IR is $-w^{\prime}(z=1)$
vs. $\mu\tilde{\mu}/2(3-q^2)^2$ (left). The duality 
in the UV is $-\tilde{w}^{\prime}(z=0)$ vs. $\mu\tilde{\mu}$ (right). 
Both are straight lines,
which verifies the duality conditions eq.(\ref{IR-duality100}) and eq.(\ref{UV-duality100}).}
\label{UV-IR-duality}
\end{figure}

We solve the Riccati equation (\ref{w-EOM1}) numerically using the adaptive algorithm
with the chemical potential as a parameter. First we adjust $\tilde{\mu}$ to obtain 
the required UV behavior of the dual solution eq.(\ref{UV-boundary2}): $\tilde{w}$ is a straight line 
going through the origin eq.(\ref{UV-boundary2}).
It corresponds to switching off the source $\tilde{a}^{(1)}=0$ on the magnetic side. 
Then, solving Riccati equation numerically,
we adjust $\mu$ to obtain
the required UV behavior of the solution on electric side eq.(\ref{UV-boundary1}): 
$A$ is a straight line going through the origin eq.(\ref{UVduality300}),
that corresponds to switching off the source $a^{(0)}=0$ on the electric side.
This step is done provided that a constant $w_1$ in the IR boundary condition 
satisfies the duality relation eq.(\ref{IR-duality100}). The gauge fields $A$ are obtained from the solutions
of the Riccati equation $w$ imposing the IR boundary conditions that provides proper normalization: 
$A(z=1)=1$ on electric side, $\tilde{A}(z=1)$
satisfies the duality condition.   
We find that in the Schwarzschild metric,
we satisfy the required UV boundary conditions for the following chemical potentials:
\begin{eqnarray}
&& {\rm E\;side:}\; \mu\geq 3.656,\label{bce2}\\
&& {\rm B\;side:}\; \tilde{\mu}\approx 1.05 \label{bcb2}
\label{}
\end{eqnarray}

\begin{figure}[!ht]
\includegraphics[width=0.35\textwidth]{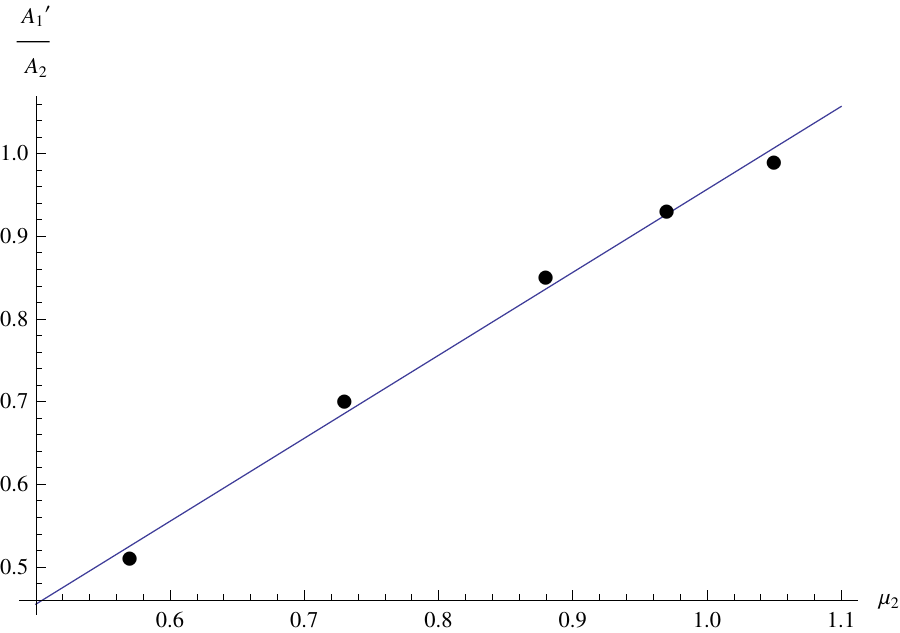}
\includegraphics[width=0.35\textwidth]{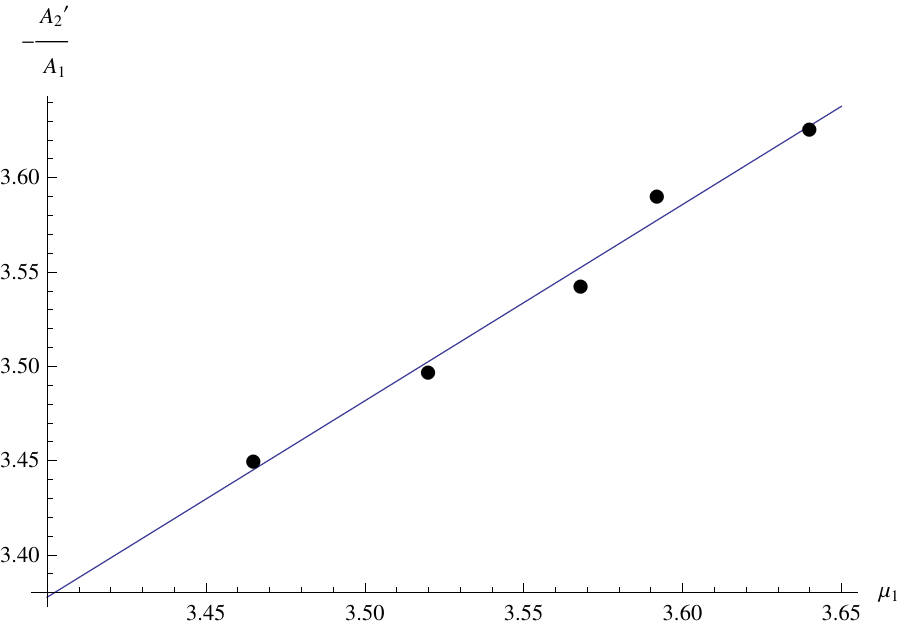}
\caption{The duality relation for the RN-AdS BH metric in the UV asymptotics with the charge increased from $q=0$ to $q=1.5$:
$A^{\prime}(z=0)/\tilde{A}(z=0)$ vs. $\tilde{\mu}$ (left), $-\tilde{A}^{\prime}(z=0)/A(z=0)$ vs. $\mu$ (right). 
Linear behavior confirms the duality conditions for the gauge fields eqs.(\ref{AdualityUV}).}
\label{Duality_for_A}
\end{figure}

The two solutions of the Riccati equations have the UV asymptotic behavior $w\sim 1/z$ ("threshold" solution) and $w\sim z$ ("new" solution)
as depicted in Fig.(\ref{Riccati_solution}). They correspond to the gauge field solutions of the equations of motion with UV behavior
$A\sim z$ (the "threshold" solution) - Dirichlet b.c. 
and $A\rightarrow {\rm const}$ (the "new" solution) - Neumann b.c. shown in Fig.(\ref{EOM_solution}).

\begin{figure}[!ht]
\includegraphics[width=0.35\textwidth]{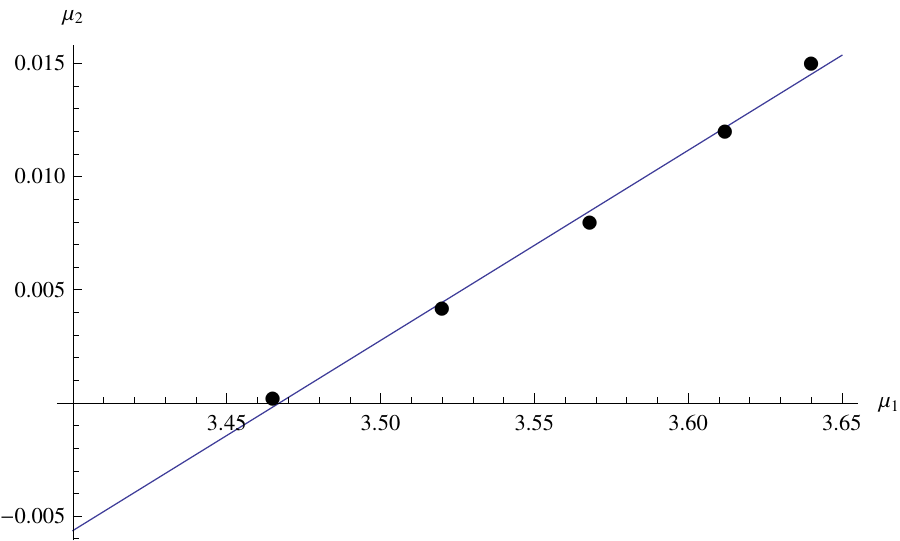}
\caption{The mapping between the chemical potentials, $\tilde{\mu}$ vs. $\mu$, as the charge is increased 
from $q=0$ to $q=1.5$.}
\label{mu_mapping}
\end{figure}
Using the adaptive algorithm for a range of AdS-RN BH charges
we find the mapping
between the two chemical potentials $\tilde{\mu}(\mu)$, Fig.(\ref{mu_mapping}). 
In other words, we find that 
for each solution of Riccati equation $w=A_x^{1\prime}/A_x^1$ at $\mu$
with Dirichlet b.c. there exist a solution
of the Riccati equation $w=\tilde{A}_y^{2\prime}/\tilde{A}_y^2$ at $\tilde{\mu}$ with Neumann b.c.. 
These two solutions are related by the duality equations (\ref{duality-condition1},\ref{duality-condition2}) and there
is a duality mapping $\mu\rightarrow \tilde{\mu}$ described by a function $\mu(\tilde{\mu})$. The $S$-duality transformation 
$E\xrightarrow{S}B$ acts on the gauge field solution as
\begin{eqnarray}
&& {\rm "electric"\, side} \xrightarrow{S} {\rm "magnetic" \, side}\\
&& {\rm large}\, \mu \xrightarrow{S} {\rm small}\, \mu \\
&& {\rm Dirichlet\, b.c.} \xrightarrow{S} {\rm Neumann\, b.c.}\\
&& {\rm source} \xrightarrow{S} {\rm v.e.v.}
\label{}
\end{eqnarray} 
and on the superfluid density and the conductivity as
\begin{eqnarray}
&& n_s/\mu \xrightarrow{S} \frac{1}{n_s/\mu}\\
&& \sigma \xrightarrow{S} \frac{1}{\sigma}\\
&& {\rm poles}\, \sigma \xrightarrow{S} {\rm zeros}\, \frac{1}{\sigma}\\
&& {\rm metal} \xrightarrow{S} {\rm insulator}
\label{}
\end{eqnarray}
and the reverse transformation "$\xleftarrow{S}$" is also true. 
It was shown for abelian $U(1)$ gauge fields in \cite{Witten} and for non-abelian gauge fields
in \cite{Rangamani}, that the generator $S$ of $SL(2,Z)$ exchanges electric and magnetic fields. 
Further, the "electric" side with $B=0$ b.c. for the gauge fields
corresponds to the Dirichlet b.c. (standard quantization), vector potential $A$ vanishing at the boundary. 
While the "magnetic" side with $E=0$ b.c. corresponds to the Neumann b.c. (alternative quantization), boundary values
of $A$ remaining unrestricted. This means that swapping between the Dirichlet and the Neumann b.c. leads to swapping 
the identification of the source and the v.e.v. \cite{McGreevy}.   

The transformation $\xrightarrow{S}$ corresponds to a superconductor - insulator
quantum critical point of bosons in $(2+1)$ dimensions \cite{Conductivity2}. 
Thus we have demonstrated that the duality relations eqs.(\ref{duality-condition1},\ref{duality-condition2}) 
are satisfied for the AdS-RN and Schwarzschild black holes  
in the bulk for all $z$, see Fig.(\ref{Duality_check}),
and in the IR/UV asymptotics, Fig.(\ref{UV-IR-duality}). We checked the duality condition
for the original gauge fields $A, \tilde{A}$ in the UV, Fig.(\ref{Duality_for_A}), while the duality
relation in the IR serves as boundary conditions to normalize the solutions.  
Therefore we can state that S-duality acts on the Riccati equation and its physically relevant solutions in a known way.

\section{On Flavor S-duality in holographic QCD with isotopic and baryonic chemical potentials}

Let us briefly discuss the possible counterpart of the S-duality for p-wave superconductor
in holographic QCD. First, let us remind what kind of phenomena are relevant for this issue.
For low-energy QCD the flavor group $U(N_f)_L \times U(N_f)_R$ plays the role of the gauge group 
in the holographic dual which is broken to the diagonal one by the chiral condensate. To fit
with the previous discussion we restrict ourselves by $N_f=2$. The abelian $U(1)_B,U(1)_A$ factors
correspond to the baryon charge and axial singlet symmetry broken by the anomaly. 
The most popular holographic models for QCD are the Sakai-Sugimoto models and D3/D7 models 
which can be thought of as the Chiral Lagrangian supplemented by the infinite tower 
of the massive vector mesons. 

In the chirally broken phase the symmetry involves global $U(1)_B\times SU(2)_I$. We would
like to consider dopped system and introduce the isotopic and baryonic chemical potentials
$\mu_I,\mu_B$. If $\mu_B\neq 0$ at small $\mu_I$ the pion condensate gets emerged and there
is a kind of rotation between chiral and pion condensates discussed in details in \cite{rebhan}.
The pion condensate yields the nontrivial supercurrent and the superfluid component.If 
$\mu_I$ increases the new phase with the vector condensate appears which is the analogue
of  p-wave superconductor. This $\rho$-meson condensate has been identified in the
Sakai-Sugimoto model \cite{aharony}, in a model with probe D7-branes embedded in anti-de Sitter space \cite{MartinAmmon1} 
and in D3/D5 model \cite{MartinAmmon2}. In D3/D7 model an instability towards vector meson condensate
was found above a critical isospin chemical potential \cite{kaminski}.
The physics of this
phase is quite clear - the mass of the vector meson decreases with $\mu_I$ and at some
critical value of the chemical potential it vanishes allowing the condensation. The appearance
of the vector condensate can be seen geometrically in the D3/D7 model as follows \cite{kaminski}.
At small isotopic chemical potential  the strings from D7 branes are attracted to the horizon
making it isotopically charged. However increasing the $\mu_I$ the strings rearrange 
and form 7-7 strings that is vector mesons. The vector condensate is formed from the flow
of 7-7 strings in the bulk.

One more relevant phenomena concerns the effects of the external magnetic field. The 
electromagnetic charge is included into the isotopic group as $Q=diag(2/3,-1/3)$ therefore
the magnetic field $B_{em}$ has the baryonic and isotopic components $QB = B_{bar}I + B_{iso} \tau_3$.
The magnetic field changes the critical values of the chemical potential for the
pion condensate formation. Moreover there are arguments in favor of formation of the
vector meson condensate above some critical $B_{crit}$ \cite{chernodub}. This is analogue of the
p-wave superconductivity once again however this phenomena is still questionable.

Turn now to the our conjecture concerning the flavor S-duality in low-energy QCD. Since the S-duality is expected
to be generalization of particle-vortex duality in 2+1 we have to identify what "`particle"' and "`vortex"'
do we mean in 3+1 case and in its 4+1 dual. In 4+1 dual nonabelian gauge theory there are two point-like objects with
different charges - gauge bosons and instantons. The corresponding 
global symmetry group generated by conserved currents is $U(1)_{top} \times U(1)_{el}$
where the topologically conserved current is defined as 
\begin{equation}
j_{\beta}= \epsilon_{\nu \mu \gamma \delta \beta} Tr F_{\mu \nu}F_{\gamma \delta}
\end{equation}
These abelian groups are the 4+1 dimensional counterpart of the pair of abelian
groups involved in S-duality transformation in 2+1 case.
In boundary low-energy QCD  particle and "`vortex"' are identified as charged vector mesons
and baryons which are the instantons in the flavor group in 4+1 and realize Skyrmions in 3+1 \cite{stephanov}.
There  is however some subtlety concerning the identification of the baryon as the instanton in 4+1
and there are serious arguments for its treatment not as the instanton but as the
dyonic instanton \cite{gk}. Such interpretation fits better with the effects of the
chiral condensate on the baryon state.

Let us emphasize that we consider the analogue of S-duality in QCD for the flavor
not color group and instead of  "`W-bosons"' and monopoles represented S-dual pair in color gauge group in 3+1 
we consider the vector mesons and baryons. One could wonder if such 
S-duality is natural from the stringy viewpoint. To this aim it is useful to remind 
that non-abelian 5d gauge theory in IIB picture is represented by the (p,q) 5-brane web.
In the brane language  vector gauge bosons and instantons in 5d gauge theory indeed form the S-dual pair being
represented by the F1 nd D1 strings attached to the 5-brane web. In the IIA picture 
the vector mesons and baryons are presented more asymmetrically.

Remark that in the
4+1 dual theory there is natural "`flux attachment"' procedure via the 
5d Witten effect. Indeed due to the 5d Chern-Simons term ($N_f>2$) at level $N_c$
the state with the topological charge $Q_{top}$ gets the electric charge
\begin{equation}
Q_e= N_c Q_{top}
\end{equation}
where $N_c$ is the rank of the color group. This is familiar picture for the
strings attachment to the baryonic vertex in the bulk \cite{witten2}.

Hence our candidates for the S-dual pair are the $(\rho, baryon)$ and therefore 
we can assume that $(\mu_I,\mu_B)$ play the same role as the $(\mu, \tilde{\mu})$
in the rest of the paper.  What kind of condensates are available? As was mentioned
at large enough $\mu_I$ the "`electric"' p-wave condensate gets formed. The same
p-wave condensate presumably gets formed at large magnetic field. In this respect the
situation resembles the QHE case if we treat $(\mu_I, B)$ as the dual pair. More interestingly
if we consider $(\mu_I,\mu_B)$ as the dual pair we have nontrivial phase structure 
in this plane when the pion condensate gets substituted by the condensate of the
vector mesons. At large enough $\mu_B$ the pion condensate gets formed hence 
we could attempt to treat $<\partial_{i}\pi>$ as the pseudovector dual required
by the S-duality.

Could we present more arguments supporting duality between QCD with isotopic 
and baryonic chemical potentials? A bit surprisingly the orbifold equivalence
provides some supporting evidences \cite{orbi1,orbi2,orbi3}. The orbifold equivalence 
relates the theories with the different groups supplemented to the projection to
the particular sector of the theory. It is well established in the perturbation theory
while its status at the nonperturbative level is more subtle. Nevertheless assuming
its validity at the nonperturbative level it was argued in \cite{orbi1,orbi2,orbi3} that
there is the large N duality between QCD with isotopic chemical potential and QCD with baryonic
chemical potential at some regions in the $(\mu_I, \mu_B)$ plane. Our interpretation
of these orbifold equivalence as the version of S-duality is new. We could also speculate
that the experimentally observed equal intercepts of meson and baryon Regge trajectories
could be related to our duality  conjecture. Usually this observation is treated 
differently as the consequence of the large diquark component inside the baryon.

Our consideration suggests that there could be the relation between the superfluid/superconducting densities
in the dual descriptions that is the superconducting components of densities 
for isotopic and the baryonic charge carriers in the
chirally broken phase. Their product presumably can be proportional to $\mu_I \mu_B$
\begin{equation}
n^s_I n^s_B \sim \mu_I \mu_B.
\end{equation}
The special attention is to large $\mu_B$ limit when the chiral symmetry is restored
and QCD is in the color-flavor locking phase supporting the color superconductivity.
This issue deserves further investigation and we hope to discuss it elsewhere.

Completing this Section remark that the S-duality followed from 5d dual which we 
conjecture as flavor S-duality in low-energy QCD could have clear-cut counterpart 
in 1+1 case. Indeed we could start with the 2+1 gauge dual involving the YM and
Chern-Simons terms. In the bulk we can define the electric U(1) and topologically
conserved U(1) current 
\begin{equation}
j_{\mu}= \epsilon_{\mu,\nu\alpha} F_{\nu \alpha}
\end{equation}
We have gauge bosons and instanton particles  in the bulk theory which carry the topological charge. They form
the S-dual pair in the bulk. In the boundary theory the instantons become the Skyrmions.
The boundary S-duality presumably exchanges the 1+1 theory with chemical potentials 
for vector mesons and for Skyrmions.

\section{Conclusions}

Recently there has been a considerable effort made to establish duality in $(2+1)$D theories  
\cite{MetlitskiVishwanath}-\cite{SeibergSenthilWangWitten}. In \cite{MetlitskiVishwanath}, a particle-vortex
duality used to study bosonic systems has been proposed to find duality for Dirac fermions in $(2+1)$
dimensions. There is also an alternative description of this duality which is accessed via an electro-magnetic duality
in the topological superconductor.  
In \cite{KarchTong},\cite{SeibergSenthilWangWitten},
a whole class of $(2+1)$D dualities is derived from an elemental duality between a bosonic field theory 
and a fermionic field theory, where the duality between the two fermion theories appears as one particular case.
There is also a connection with duality between $(2+1)$D Chern-Simons theories \cite{Aharony}.
Duality for the Dirac fermion in $(2+1)$D relates fermionic theories at nonzero density and in the magnetic field, that
gives a powerful tool to study the fractional quantum Hall effect, in particular particle-hole symmetric quantum
Hall states \cite{Son}.         

In this paper we  suggest a duality between fermion theories in (2+1)D where the charge density and the magnetic field are interchanged on the two sides of the duality. We establish this duality via electro-magnetic duality in a 
holographic dual theory in (3+1)D.  
Specifically, we suggested some version of generalization of S-duality in the holographic theories
involving non-abelian $SU(2)$ gauge field in the bulk. 
This permits to consider a class of theories which include the condensates and exhibit phase transitions.
Using the p-wave holographic superconductor we construct the "electric" and "magnetic" sides of the EM duality, where the holographic
gauge fields are solutions of the EOM and they are related by the $S$ duality transformation. 
As in the abelian case, for the non-abelian gauge fields
the generator $S$ exchanges the electric and magnetic fields. 
"Electric" side has $B=0$ b.c. for the gauge fields which are equivalent to Dirichlet b.c.,
where the vector potential vanishes on the boundary. 
While "magnetic" side has $E=0$ b.c. which are equivalent to Neumann b.c., that leave the boundary value
of the vector potential unrestricted. Action of the $S$ transformation leads to the exchange between the source and v.e.v. in the solution. 

We found analytically the gauge field components and asymptotics of the solutions. 
While we solved numerically for the solutions in the $AdS$ bulk on both sides of the duality.
"Electric" side is characterized by a standard $p$-wave superconductor solution 
which persists from a critical chemical potentials $\mu_c\approx 3.656$ - a "threshold" solution. 
"Magnetic" side exhibits a new type of instability 
for very small chemical potentials  - a "new" superconductor solution.
"Neutral" superconductor was found in different AdS gravitational theories in 
\cite{NewInstability1,NewInstability2,NewInstability3,NewInstability4}, but always
at small $\mu$ and with Neumann b.c. 
Assuming the EM-duality condition between the bulk solutions, 
we obtain the mapping between the chemical potentials
of the "threshold" and "new" solutions 
which is a functional dependence $\mu$ vs. $\tilde{\mu}$. 
It would be interesting to extend the S-duality transformation by the
T-duality to get the $SL(2,Z)$ group.

We find the indication for S-duality relation between superfluid densities which is similar 
to the relation for conductivities. Specifically we observe that the dimensionless ratio
$n_s/\mu$ and its $S$-dual CFT pair $\tilde{n}_s/\tilde{\mu}$
are the inverse to each other $n_s/\mu \cdot \tilde{n}_s/\tilde{\mu} \sim 1$.
The relation for conductivities is 
$\sigma\tilde{\sigma}\sim 1$.
It is consistent with the fact 
that $S$-duality corresponds to the metal - insulator transition \cite{Conductivity2}.

We made conjecture on the possible S-duality in the SU(2) flavor sector in large $N_c$ QCD based on the holographic 
picture. At the holographic side S-duality maps electrically and topologically charged states
in 4+1 bulk nonabelian gauge  theory which substitute the electric and magnetic frames in 3+1. Physically at the boundary
the topological sector corresponds to baryons while the electric sector to the vector mesons.
The corresponding chemical potentials are isotopic and baryonic ones and the large isotopic 
chemical potential yields the p-wave superconductor for the chirally broken phase. 
On the other hand the baryonic chemical potential yields 
pion condensate in the chirally broken phase  and the Cooper condensates
in the color-flavor locking phase. Therefore it would be interesting to explore
further the interplay between the isotopic and baryonic sectors at chirally broken
phase treated as the version of S-duality  and on the other hand the interplay between the chirally broken phase
supplemented by magnetic field and the phase with the color superconductivity.

\section*{Acknowledgments}

The authors thank Martin Ammon, Emil Akhmedov, Gerald Dunne, Johanna Erdmenger, Alex Kovner, Max Metlitski, Nikita Nekrasov, 
Andreas Schmitt, Dam Son, David Tong, Ashvin Vishwanath and William Witczak-Krempa for discussions. 
Special thanks are to Andy O'Bannon for valuable contributions to the work.  
This work is supported, in part, by grants 
RFBR 15-02-02092 and Russian Science Foundation grant for IITP 14-50-00150 (A.G.), and
RFBR 16-02-01021 (A.Z.).

\clearpage
\appendix

\section{SL(2,Z) invariance of the axio-dilaton SU(2) gauge action}\label{appendix:1}

The Einstein-Maxwell action coupled to an axion and a dilaton fields in $3+1$ dimensions 
has been considered in \cite{Burgess},\cite{Gibbons}. We generalize their action
to the case of SU(2) non-abelian gauge fields.
The gauge-gravity action coupled to an axio-dilaton is given by
\begin{eqnarray}
S_{\phi,\chi} &=& -\int d^4x \sqrt{-g}\left(\frac{1}{2\kappa^2}
\left[R-2\Lambda +\frac{1}{2}(\partial_{\mu}\phi\partial^{\mu}\phi
+{\rm e}^{2\phi}\partial_{\mu}\chi\partial^{\mu}\chi)
\right]\right.\nonumber\\
  && \left.+\frac{1}{4}{\rm e}^{-\phi}F_{\mu\nu}F^{\mu\nu} - \frac{1}{4}\chi F_{\mu\nu}\ast F^{\mu\nu}
\right)
\label{action}
\end{eqnarray}
where $F_{\mu\nu}=\partial_{\mu}A_{\nu}-\partial_{\nu}A_{\mu}+[A_{\mu},A_{\nu}]$, with non-abelian gauge field
$A_{\mu}=A_{\mu}^a \tau_a$ and $\tau_a$ are generators of the $SU(2)$ group, $[\tau^a,\tau^b]=i\epsilon^{abc}\tau^c$,
related to the Pauli matrices by $\tau_a=\sigma_a/2i$.
We introduce the covariant derivative
$D_{\mu}=I\partial_{\mu}-ig\tau^aA_{\mu}^a$, therefore field strength is
$[D_{\mu},D_{\nu}]=-ig\tau^aF_{\mu\nu}^a$ where the coupling $g$ is defined further.
The dual field strength is obtained by applying 
the Hodge star operation  
$\ast F_{\mu\nu}:=\frac{1}{2}\epsilon_{\mu\nu\lambda\rho}F^{\lambda\rho}$,
where completely antisymmetric Levi-Civita tensor $\epsilon_{\mu\nu\rho\lambda}$ 
has a factor of $\sqrt{-g}$ with $g={\rm det}g_{\mu\nu}$ extracted and transforms as a tensor and not as a tensor density,
and indexes are freely raised and lowered using the metric $g_{\mu\nu}$ whose signature is Lorentzian
$(-\;+\;+\;+)$. Since we are working in four dimensions transforming two-form $F_{\mu\nu}$,
the Hodge dual applied twice results $\ast\ast=s=-1$ where $s$ is the signature of inner product on the manifold.
In eq.(\ref{action}), two scalar fields are dilaton $\phi$ and axion $\chi$.
The constant $\Lambda=3/L^2$ is the AdS cosmological constant
and $\kappa^2=8\pi G$ is the Newton's constant. The weak curvature means $\kappa^2/L^2 \ll 1$.
The relation to the gauge coupling $g$ and the theta-angle $\theta$ is 
\begin{equation}
{\rm e}^{-\phi}=\frac{1}{g_E^2}=\frac{1}{g^2},\;\;\;
\chi=\frac{1}{g_B^2}=\theta,
\end{equation}
where subscripts $E$ and $B$ stand for the electric and magnetic part.
Therefore weak coupling corresponds to ${\rm e}^{\phi}\ll 1$.

The matter fields, dilaton and axion, provide couplings of the gauge field action. They are added to ensure the existence
of a duality group. The SL(2,Z) duality of Maxwell U(1) gauge fields in gravity-dual four dimensions 
has been shown by Witten in \cite{Witten}. He also suggested
that generalization to a non-abelian gauge fields is possible when "special collections of matter fields are included".

As in \cite{Burgess}, we define the axio-dilaton by a complex variable
\begin{equation}
\tau:=\chi+i{\rm e}^{-\phi}=\theta+\frac{i}{g^2},
\end{equation}
(not to confuse with the SU(2) generator),
and the axio-dilaton action is rewritten
\begin{equation}
L_{\phi,\chi}\sim \partial_\mu\phi\partial^\mu\phi+{\rm e}^{2\phi}\partial_\mu\chi\partial^\mu\chi
=\frac{\partial_{\mu}\tau\partial^{\mu}\bar\tau}{({\rm Im}\tau)^2}
\label{axio-dilaton}
\end{equation}
We define matrix of $SL(2,R)$ transformation
\begin{equation}
M=
\left(\begin{array}{cc}
p & q\\
r & s
\end{array}
\right)\in {\rm SL(2,R) }
\label{matrix_S11}
\end{equation}
where 
$p,\;q,\;r,\;s$ are real numbers and ${\rm det}M=1$, that is $sp-qr =1$.
The SL(2,R) transformation acts on a axio-dilaton and on the metric as \cite{Burgess}
\begin{equation}
\tau\rightarrow \tilde{\tau}= \frac{p\tau+q}{r\tau+s},\;\;\;
g_{\mu\nu}\rightarrow \tilde{g}_{\mu\nu} =g_{\mu\nu},
\label{transformation_tau}
\end{equation}
where tilde distinguishes the variables which undergone the SL(2, R) transformation.
We rewrite eq.(\ref{transformation_tau}) 
\begin{equation}
\tau \rightarrow \frac{pr\|\tau\|^2+qs+ps2{\rm Re}\tau-\bar{\tau}}{\|r\tau+s\|^2}
\label{transformation_tau2}
\end{equation}
where we used $sp-qr=1$.  
Therefore, 
\begin{equation}
\partial_{\mu}\tau\partial^{\mu}\bar\tau\rightarrow 
\frac{\partial_{\mu}\tau\partial^{\mu}\bar\tau}{\|r\tau+s\|^4},\;\;
{\rm Im}\tau\rightarrow \frac{{\rm Im}\tau}{\|r\tau+s\|^2}
\end{equation}
and the axio-dilaton action $S_{\phi,\chi}$ eq.(\ref{axio-dilaton}) is $SL(2,R)$ invariant.
To see the $SL(2,R)$ action on the gauge field we define, as in  \cite{Gibbons},\cite{Burgess}
\begin{equation}
G^{\mu\nu}:=-\frac{2}{\sqrt{-g}}\frac{\delta S}{\delta F_{\mu\nu}}
\end{equation}
and obtain
\begin{eqnarray}
G^{\mu\nu} &=& {\rm e}^{-\phi}F^{\mu\nu}-\chi\ast F^{\mu\nu}.
\label{G}
\end{eqnarray}
Written in terms of complex quantities \cite{Burgess}
\begin{eqnarray}
{\mathcal F}_{\mu\nu}&:=& F_{\mu\nu} + i\ast F_{\mu\nu},\\
{\mathcal G}_{\mu\nu}&:=& \ast G_{\mu\nu} - iG_{\mu\nu},
\label{complex}
\end{eqnarray}
eq.(\ref{G}) takes a compact form
\begin{equation}
{\mathcal G}^{\mu\nu}=\bar{\tau} {\mathcal F}^{\mu\nu}.
\label{relation}
\end{equation} 
The $SL(2,Z)$ acts on the gauge fields as \cite{Burgess}
\begin{eqnarray}
\left(\begin{array}{c}
{\mathcal G}_{\mu\nu}\\
{\mathcal F}_{\mu\nu}
\end{array}\right)\rightarrow 
\left(\begin{array}{c}
\tilde{\mathcal G}_{\mu\nu}\\
\tilde{\mathcal F}_{\mu\nu}
\end{array}\right)
=\left(\begin{array}{cc}
p&q\\
r&s
\end{array}
\right)
 \left(\begin{array}{c}
{\mathcal G}_{\mu\nu}\\
{\mathcal F}_{\mu\nu}
\end{array}\right),
\label{transformation}
\end{eqnarray}
where tilde denotes the transformed variables.
Relation eq.(\ref{relation}) in invariant under SL(2,R). Indeed, after transformation we have
\begin{equation}
p{\mathcal G}+q{\mathcal F} =\frac{p\bar{\tau}+q}{r\bar{\tau}+s}(r{\mathcal G}+s{\mathcal F})
\end{equation}
which is reduced to eq.(\ref{relation}) provided $sp-qr=1$.
From eq.(\ref{transformation}) follows SL(2,R) transformation for the field strengths 
\begin{eqnarray}
F_{\mu\nu}^a &\rightarrow & \tilde{F}_{\mu\nu}^a = s F_{\mu\nu}^a +r\ast G_{\mu\nu}^a,\label{field_strength_transform0}\\
G_{\mu\nu}^a &\rightarrow & \tilde{G}_{\mu\nu}^a = p G_{\mu\nu}^a -q\ast F_{\mu\nu}^a,
\label{field_strength_transform}
\end{eqnarray}
where we used for the double Hodge duality $\ast\ast=-1$.
Using the definition for $G$ eq.(\ref{G}) and eqs.(\ref{field_strength_transform0},\ref{field_strength_transform}),
the SL(2,R)-dual field strength $F_{\mu\nu}$ is  
\begin{equation}
F_{\mu\nu}^a\rightarrow \tilde{F}_{\mu\nu}^a = r{\rm e}^{-\phi}\ast F_{\mu\nu}^a+(s+r\chi)F_{\mu\nu}^a,
\label{transformation_F}
\end{equation}
The gauge action 
\begin{eqnarray}
L_{F,\ast F} &\sim & F_{\mu\nu}G^{\mu\nu}\rightarrow
(sF_{\mu\nu}+r\ast G_{\mu\nu})(pG^{\mu\nu}-q\ast F^{\mu\nu})
\nonumber\\
&=& (sp-qr)F_{\mu\nu}G^{\mu\nu}=F_{\mu\nu}G^{\mu\nu},
\end{eqnarray}
is invariant under SL(2,R) transformation. 
We used $\ast G_{\mu\nu}\ast F^{\mu\nu}=F_{\mu\nu}G^{\mu\nu}$ and $sp-qr=1$; and the Lagrangian satisfies
a differential constraint
$\ast G_{\mu\nu} G^{\mu\nu}=F_{\mu\nu}\ast F^{\mu\nu}$  with $G_{\mu\nu}$ given by eq.(\ref{G}) 
and $\|\tau\|^2=1$, \cite{Gibbons}.

The SL(2,R) invariance can be shown using the electric and magnetic fields. We define the electric intensity
$E_i^a=F_{i 0}^a$ and magnetic induction $B_i^a=\frac{1}{2}\epsilon_{ijk}F^{jka}$, with $\ast {\bf E} = -{\bf B}$ and
$\ast {\bf B} = {\bf E}$, and the electric induction
$D_i^a=G_{i0}^a$ and magnetic intensity $H_i^a=\frac{1}{2}\epsilon_{ijk}G^{jka}$ where $G_{\mu\nu}$ 
is given by eq.(\ref{G}), 
\begin{eqnarray}
{\bf D} &=& \frac{1}{\sqrt{-g}}\frac{\partial S}{\partial {\bf E}}={\rm e}^{-\phi}{\bf E}+\chi {\bf B},\\
{\bf H} &=& -\frac{1}{\sqrt{-g}}\frac{\partial S}{\partial {\bf B}}={\rm e}^{-\phi}{\bf B} -\chi {\bf E},
\end{eqnarray} 
where ${\bf E}={\bf E}^a \tau^a$ with $\tau^a\in SU(2)$ and the same for other vector fields.  
From eq.(\ref{transformation}), SL(2,R) transformation for the fields is
\begin{eqnarray}
&&{\bf E}^a\rightarrow s{\bf E}^a-r{\bf H}^a,\;\; {\bf B}^a\rightarrow s{\bf B}^a+r{\bf D}^a,
\label{field_transform1} 
\end{eqnarray}
and
\begin{eqnarray}
&& {\bf D}^a\rightarrow p{\bf D}^a+q{\bf B}^a,\;\; {\bf H}^a\rightarrow p{\bf H}^a-q{\bf E}^a.
\label{field_transform2}
\end{eqnarray} 
Therefore the action  
\begin{eqnarray}
L_{{\bf E},{\bf B}} &\sim & {\bf E}\cdot{\bf D} + {\bf B}\cdot{\bf H}\rightarrow
(s{\bf E}-r{\bf H})(p{\bf D}+q{\bf B})+(s{\bf B}+r{\bf D})(p{\bf H}-q{\bf E}) \nonumber\\
&=& (sp-qr)({\bf E}{\bf D}+{\bf B}{\bf H})={\bf E}\cdot{\bf D}+{\bf B}\cdot{\bf H}
\end{eqnarray}
is invariant under SL(2,R).

In general, only abelian $U(1)$ gauge action and the corresponding Maxwell equations are invariant
under the $SL(2,R)$ duality \cite{Gibbons},\cite{Gibbons_em}.
Indeed, performing the $SL(2,R)$ transform eq.(\ref{field_strength_transform0},\ref{field_strength_transform}) in the
Yang-Mills equation $D\ast G=0$  
\begin{equation}
(D_{\mu}\ast G^{\mu\nu})^a=0,
\end{equation} 
and the Bianchi identity $D F=0$, 
\begin{equation}
(D_{\mu} F_{\nu \kappa})^a+(D_{\kappa} F_{\mu \nu})^a+(D_{\nu} F_{\kappa\mu})^a=0,
\end{equation} 
the problem is caused by a covariant derivative, where the $SL(2,R)$ transformation should be written
for the gauge field and not for the field strength; therefore this transformation has an integral, non-local character. 
However, there exist solutions which are $S$-invariant in all the space or in the asymptotic regions. One example
is an AdS instanton solution which is self-dual in the whole space. Further we show that the p-wave superconductor is $S$-invariant
in the UV and IR.   
Equations of motion for axion and dilaton are SL(2,R) invariant.

Also, one should have in mind the following remark. 
In string theory, $SL(2,R)$ group generally holds 
in the classical approximation, and is broken down to a discrete subgroup by
the quantum effects \cite{Burgess}. Specifically,   
the symmetry is broken by the presence of objects whose charges are quantized,
for example, the $(m, n)$-string breaks $SL(2,R)$ down to $SL(2,Z)$ \cite{Burgess}.
In what follows we consider $SL(2,Z)$ for our applications.

$SL(2,Z)$ is generated by the two matrices \cite{Witten}, $S$ is the analog of electric-magnetic duality
\begin{eqnarray}
S=\left(\begin{array}{cc}
0&1\\
-1&0
\end{array}\right),
\end{eqnarray}
and $T$ acts on a topological term $\theta\int d^4x F\ast F$ by shifting $\theta\rightarrow \theta+2\pi$
\begin{eqnarray}
T=\left(\begin{array}{cc}
1&1\\
0&1
\end{array}\right).
\end{eqnarray}
The $S^2=-1$ is the central element, and $(ST)^3=1$, while $S$ and $T$ do not commute. 
The action of $S$ and $T$ transformation on the boundary conformal field theories and the relation to the AdS 
gravitational theories has been discussed in \cite{Witten}. On the $S$-operation, gravitational
theory in AdS space has two
different CFT duals on the boundary, depending on which boundary condition one chooses
to impose. It has been shown for the abelian $U(1)$ gauge fields \cite{Witten},
that the generator $S$ of $SL(2,Z)$ exchanges electric and magnetic fields, and it corresponds
from the AdS point of view to replacing the boundary condition ${\bf B} = 0$ (generally where $B$ is specified) 
which is the "electric" side
with ${\bf E} = 0$ (generally where $E$ is specified) which is the "magnetic" side of electric-magnetic duality. 
Further \cite{Witten}, ${\bf B} = 0$ boundary conditions for the gauge fields  are the analogs 
to Dirichlet boundary conditions � they say that vector potential {\bf A} vanishes on the
boundary, up to a gauge transformation. While ${\bf E} = 0$ boundary conditions are analogous to free
or Neumann boundary conditions. They leave the boundary values of {\bf A} unrestricted.
We will use this fact later to identify the source and the v.e.v. in the asymptotic behavior of gauge fields 
at the AdS boundary.

The $S$-operator acts on the axio-dilaton eq.(\ref{transformation_tau2}) as
\begin{equation}
\tau\rightarrow \tilde{\tau} = -\frac{1}{\tau},
\label{Sdual_tau}
\end{equation}
where we used $p=s=0$ and $q=-r=1$ in the SL(2,R) matrix eq.(\ref{matrix_S11}). 
While the $S$-duality transformation is interesting to perform on the full gauge and axio-dilaton action
eq.(\ref{action}), we restrict ourselves to the vanishing axion field
\begin{equation}
\chi=0.
\end{equation} 
According to eq.(\ref{Sdual_tau}) and definition of $\tau=\chi+i{\rm e}^{-\phi}$, 
the axion field is not generated by the S-duality transformation.
In this case 
the S-operator acts on the gauge field strength eq.(\ref{field_strength_transform0}) as
\begin{equation}
F_{\mu\nu}^a\rightarrow \tilde{F}_{\mu\nu}^a = -\ast G_{\mu\nu}^a.
\label{Sdual_FG}
\end{equation}
or written explicitly,
\begin{equation}
F_{\mu\nu}^a\rightarrow \tilde{F}_{\mu\nu}^a = -{\rm e}^{-\phi}\ast F_{\mu\nu}^a.
\label{Sdual_F}
\end{equation}
Eqs.(\ref{Sdual_tau},\ref{Sdual_F}) express a familiar electric-magnetic duality
where the field strength transforms into a Hodge-dual one and the coupling transformation is $g^2\rightarrow \frac{1}{g^2}$, 
therefore the weak-strong coupling regimes are interchanged. In $(3+1)$-d the Hodge-dual is defined
as $\star F = \frac{\sqrt{-g}}{4}\epsilon_{\mu\nu\rho\sigma}F^{\rho\sigma}dx^{\mu}\wedge dx^{\nu}$.

\section{p-wave superconductor in $AdS_5$}\label{app:a}

We write the Riccati equation for the $AdS_5$ p-wave superconductor and 
obtain the two solutions imposing Dirichlet and Neumann boundary conditions.
However, these solutions are not related to each other by the EM duality.
We discuss $AdS_5$ case here, because there is a known p-wave superconducting solution for it,
and we can test our numerical solution again the analytical one.
As we discuss in the main text,
the equations of motion for the "condensate" components on E- and B-sides of duality transformation,
$A_x^{1}$ and $A_y^{2}$ respectively, are the same. Also we substitute for $A_t^3=\mu(1-z^2)$ and $\tilde{A}_t^3=\tilde{\mu}(1-z^2)$.
In what follows we use one letter for both gauge components
$A_x^{1},A_y^{2}\rightarrow A$, and one $\mu$. In the Schwarzschild metric with the $AdS_5$ asymptotic behavior, the EOM reads
\begin{eqnarray}
&& A^{\prime\prime}+\left(\frac{f^{\prime}}{f}+\frac{1}{z}\right)A^{\prime}+\frac{\mu^2(1-z^2)^2}{z^4f^2}A=0,\label{eom_ads5}\\
&& f(z) = \frac{1}{z^2}-z^2,
\label{}
\end{eqnarray}    
and explicitly it is
\begin{eqnarray}
&& A^{\prime\prime}-\frac{1+3z^4}{z(1-z^4)}A^{\prime}+\frac{\mu^2}{(1+z^2)^2}A=0\\
&& A(z=1)=const,\\
&& A^{\prime}(z=1)=0, 
\label{}
\end{eqnarray} 
where we specified the IR boundary conditions, which mean regularity. 

\begin{figure}[!ht]
\includegraphics[width=0.35\textwidth]{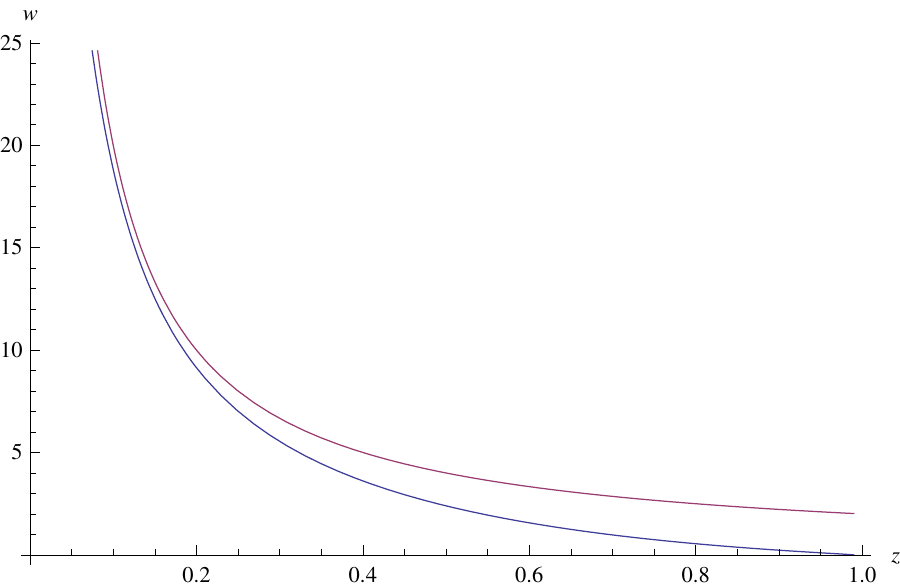}
\includegraphics[width=0.35\textwidth]{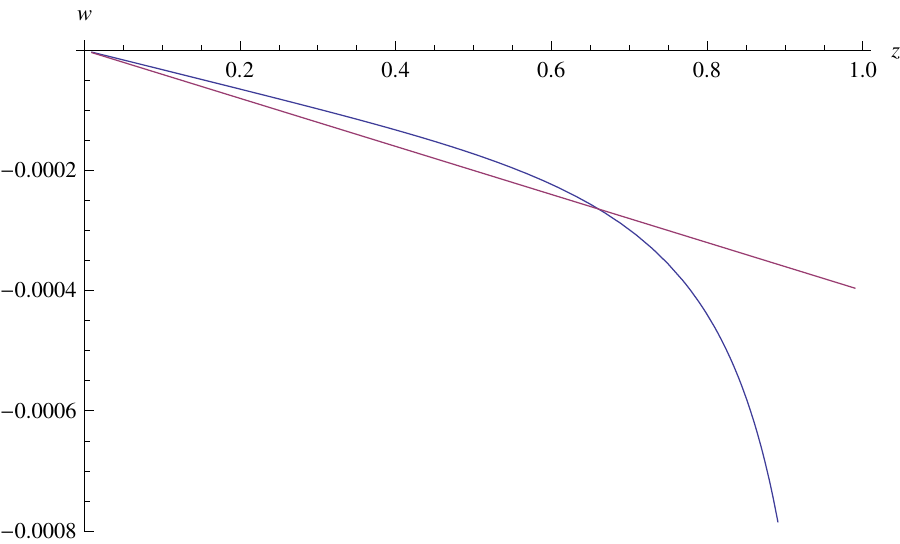}
\caption{Two solutions $w$ of the Riccati equation in the $AdS_5$: a standard "threshold" solution for $\mu=4$, $w_1=0$ (left) and a new "new" solution
for $\mu=0.001$, $w_1=-0.009$ (right). Upper curve in the right plot is $w=2/z$. Straight line in the left plot is $w=-0.0004 z$.}
\label{Riccati_solutionAdS5}
\end{figure}

In the UV $z=0$ the asymptotic behavior of the gauge field is
\begin{equation}
A=A^{(0)}+A^{(1)}z^2+....
\end{equation} 
In the UV the Dirichlet and Neumann boundary conditions are given in the main text,
that dictate one UV boundary condition for the Riccati equation. The Riccati equation corresponding to eq.(\ref{eom_ads5}) 
for the variable $w=A^{\prime}/A$ reads
\begin{eqnarray}
&& w^{\prime}-\frac{1+3z^4}{z(1-z^4)}A^{\prime}+\frac{\mu^2}{(1+z^2)^2}A=0 \label{Riccati_equationAdS5}\\
&& w(z=1)=w_1,
\label{}
\end{eqnarray}
where we impose regularity at the IR by taking $w_1$ is a constant.

\begin{figure}[!ht]
\includegraphics[width=0.35\textwidth]{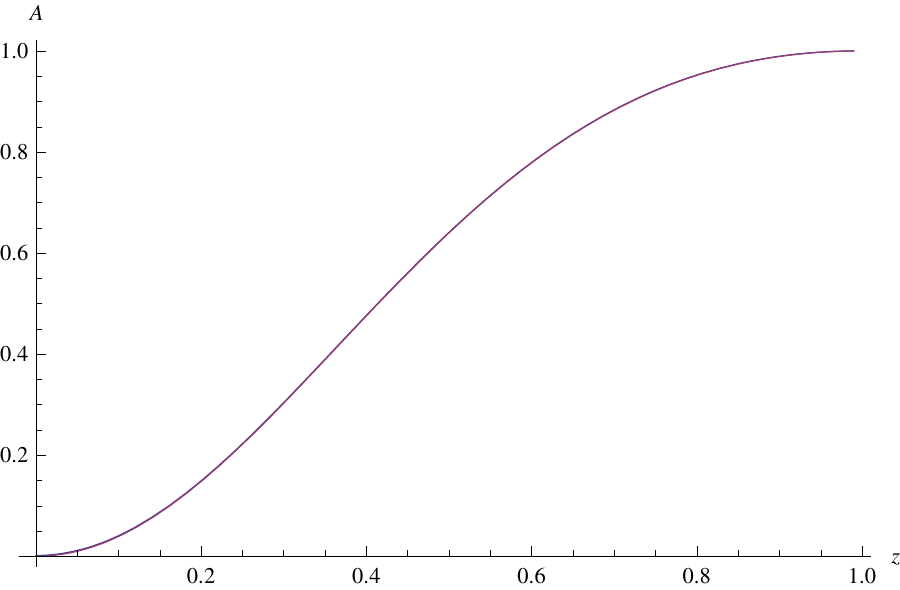}
\includegraphics[width=0.35\textwidth]{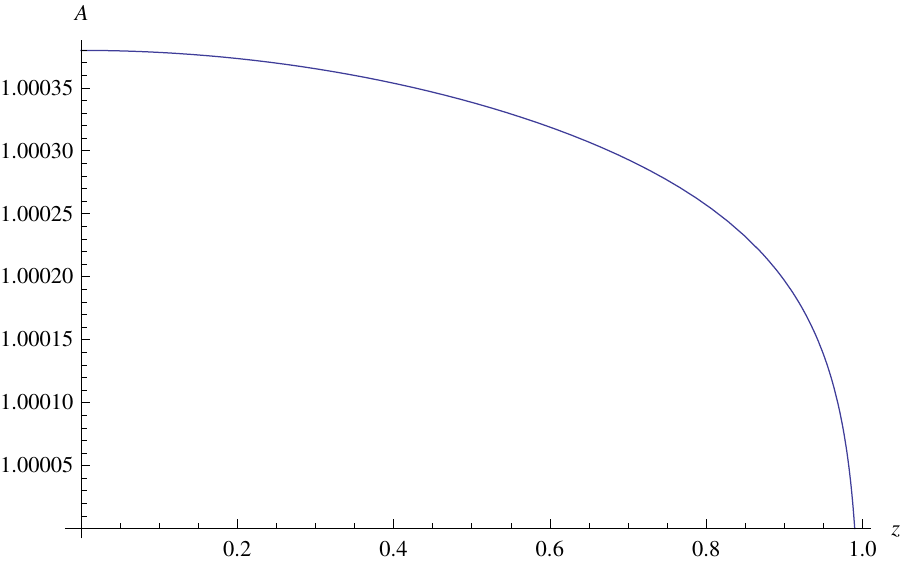}
\caption{Two solutions for the gauge field $A$ of the equation of motion in the $AdS_5$: a standard "threshold" solution for $\mu=4$, $w_1=0$ (left) and a new "new" solution for $\mu=0.001$, $w_1=-0.009$ (right). Two coinciding curves for numerical and analytical solutions $4z^2/(1+z^2)^2$ are shown in the left panel.}
\label{EOM_solutionAdS5}
\end{figure}
 
We vary two parameters $\mu$ and $w_1$ to obtain the needed UV behavior for the Dirichlet and Neumann boundary conditions
\begin{eqnarray}
&& {\rm Dirichlet:}\; w(z=0)=\frac{2}{z}\rightarrow \infty,\label{bce11}\\
&& {\rm Neumann:}\; w(z=0)=0.\label{bcb11}
\label{}
\end{eqnarray}   
As noted before, the analytic solution for the p-wave superconductor satisfying the Dirichlet boundary condition is known to be \cite{AdS5}
\begin{eqnarray}
A &=& \frac{4z^2}{(1+z^2)^2},\\
\mu_c &=& 4,
\label{}
\end{eqnarray}
where $\mu_c$ is the critical chemical potential. 

We solve the Riccati equation \ref{Riccati_equationAdS5} in $AdS_5$ numerically, and obtain the two solutions: a "threshold" solution for $\mu=4$
and a "new" solution for $\mu=0.001$ Fig.(\ref{Riccati_solutionAdS5}), which satisfy the Dirichlet and Neumann b.c. eqs.(\ref{bce11},\ref{bcb11}), respectively.
The corresponding solutions for the gauge fields are depicted in Fig.(\ref{EOM_solutionAdS5}).


\begin{thebibliography}{99}

\bibitem{Witten}
E. Witten, "SL(2,Z) Action On Three-Dimensional Conformal Field Theories With Abelian
Symmetry", in Shifman, M. (ed.) et al.: From fields to strings, vol. 2, 1173-1200,
[hep-th/0307041].\\
     
\bibitem{kapustin}
A. Kapustin and M. Strassler, "On Mirror Symmetry In Three Dimensional Abelian
Gauge Theories", [hep-th/9902033].\\

\bibitem{Gibbons}
G. W. Gibbons and D. A. Rasheed, "SL(2,R) Invariance of Non-Linear Electrodynamics
Coupled to An Axion and a Dilaton", Phys. Lett. {\bf B} 365 (1996) 46 [hep-th/9509141].\\

\bibitem{Gibbons_em}
G W Gibbons, D A Rasheed, "Electric-Magnetic Duality Rotations in Non-Linear Electrodynamics", 
Nucl. Phys. B {\bf 454} (1995) 185 [hep-th/9506035].\\

\bibitem{Burgess}
A. Bayntun, C.P. Burgess, B. P. Dolan, S.-S. Lee, 
"AdS/QHE: Towards a Holographic Description of Quantum Hall Experiments",
New J. Phys. 13 (2011) 035012 [arXiv:1008.1917].\\

\bibitem{rene} 
  M.~Lippert, R.~Meyer and A.~Taliotis,
  ``A holographic model for the fractional quantum Hall effect,''
  JHEP 1501 (2015) 023,
  doi:10.1007/JHEP01(2015)023
  [arXiv:1409.1369 [hep-th]].\\
  
\bibitem{Gubser}
 S. S. Gubser, S. S. Pufu, "The gravity dual of a p-wave superconductor",
JHEP 0811 (2008) 033 [arXiv:0805.2960].\\

\bibitem{RobertsHartnoll}
M. M. Roberts, S. A. Hartnoll, 
"Pseudogap and time reversal breaking in a holographic superconductor",
[arXiv:0805.3898].\\

\bibitem{cai}
 Rong-Gen Cai, Song He, Li Li, and Li-Fang Li. 
"A Holographic Study on Vector
Condensate Induced by a Magnetic Field",
JHEP, 1312:036, 2013.\\
Rong-Gen Cai, Li Li, and Li-Fang Li. 
"A Holographic P-wave Superconductor Model",
JHEP, 1401:032, 2014.\\

\bibitem{AdS5}
P. Basu, J. He, A. Mukherjee, H.-H. Shieh,
"Superconductivity from D3/D7: Holographic Pion Superfluid",
JHEP 0911 (2009) 070 [arXiv:0810.3970].\\
C. P. Herzog, S. S. Pufu, "The Second Sound of SU(2)",
JHEP 0904 (2009) 126 [arXiv:0902.0409].\\
P. Basu, J.-H. Oh, "Analytic Approaches to anisotropic Holographic Superfluids",
JHEP 07 (2012) 106 [arXiv:1109.4592].\\

\bibitem{Papadimitriou}
I. Papadimitriou, A. Taliotis, "Riccati equations for holographic 2-point functions",
JHEP 04 (2014) 194 [arXiv:1312.7876].\\
 
\bibitem{Rangamani}
O. Aharony, D. Marolf, M. Rangamani,
"Conformal field theories in anti-de Sitter space",
JHEP 1102 (2011) 041 [arXiv:1011.6144].\\

\bibitem{McGreevy}
T. Faulkner, H. Liu, J. McGreevy, D. Vegh,
"Emergent quantum criticality, Fermi surfaces, and AdS2",
Phys.Rev. D {\bf 83} (2011) 125002 [arXiv:0907.2694].\\ 

\bibitem{ErdmengerHomes}
J. Erdmenger, P. Kerner, S. Muller,
"Towards a Holographic Realization of Homes' Law",
[arXiv:1206.5305].\\

\bibitem{ErdmengerRho1}
M. Ammon, J. Erdmenger, P. Kerner, M. Strydom,
"Black Hole Instability Induced by a Magnetic Field",
Phys. Lett. B {\bf 706} (2011) 94 [arXiv:1106.4551].\\ 

\bibitem{ErdmengerRho2}
Y. -Y. Bu, J. Erdmenger, J. P. Shock and M. Strydom,
"Magnetic field induced lattice ground states from holography", [arXiv:1210.6669].\\

\bibitem{Kachru_Sachdev}
N. Bao, S. Harrison, S. Kachru, S. Sachdev,
"Vortex Lattices and Crystalline Geometries",
[arXiv:1303.4390].\\

\bibitem{O'Bannon}
M. Ammon, J. Erdmenger, V. Grass, P. Kerner, A. O'Bannon,
"On Holographic p-wave Superfluids with Back-reaction",
Phys.Lett. B {\bf 686} (2010), 192 [arXiv:0912.3515].\\ 

\bibitem{Sachdev}
Subir Sachdev,
"Compressible quantum phases from conformal field theories in 2+1 dimensions",
Phys. Rev. D {\bf 86} (2012), 126003 [arXiv:1209.1637].\\

\bibitem{Faulkner_Iqbal}
T. Faulkner and N. Iqbal, "Friedel oscillations and horizon charge in 1D holographic liquids", 
[arXiv:1207.4208].\\ 

\bibitem{Mulligan_Nayak_Kachru}
Michael Mulligan, Chetan Nayak, Shamit Kachru,
"Effective Field Theory of Fractional Quantized Hall Nematics",
[arXiv:1104.0256].\\

\bibitem{Hook_Kachru_Torroba_Wang}
Anson Hook, Shamit Kachru, Gonzalo Torroba, Huajia Wang,
"Emergent Fermi surfaces, fractionalization and duality in supersymmetric QED",
[arXiv:1401.1500].\\

\bibitem{Bolognesi}
Stefano Bolognesi, David Tong,
"Magnetic Catalysis in AdS4", [arXiv:1110.5902].\\
Stefano Bolognesi, J. N. Laia, David Tong, Kenny Wong,
"A Gapless Hard Wall: Magnetic Catalysis in Bulk and Boundary",
JHEP 1207 (2012), 162 [arXiv:1204.6029].\\
  
\bibitem{Gubankova}
E. Gubankova, M. Cubrovic, J. Zaanen, "Exciton-driven quantum phase transitions in holography",
Phys. Rev. D {\bf 92}, 086004 (2015) [arXiv:1412.2373].\\

\bibitem{Schmitt}
Florian Preis, Anton Rebhan, Andreas Schmitt,
"Inverse magnetic catalysis in dense holographic matter",
JHEP 1103 (2011), 033 [arXiv:1012.4785].\\
Florian Preis, Anton Rebhan, Andreas Schmitt,
"Inverse magnetic catalysis in field theory and gauge-gravity duality",
Lect. Notes Phys. "Strongly interacting matter in magnetic fields" 
(Springer), edited by D. Kharzeev, K. Landsteiner, A. Schmitt, H.-U. Yee [arXiv:1208.0536].\\
Andreas Schmitt, "Inverse magnetic catalysis in dense holographic matter", Presentation, \href{http://hep.itp.tuwien.ac.at/~aschmitt/frankfurt.pdf}{http://hep.itp.tuwien.ac.at/~aschmitt/frankfurt.pdf}

\bibitem{Thies}
M. Thies,
"From relativistic quantum fields to condensed matter and back again: Updating the Gross-Neveu phase diagram",
J. Phys. A {\bf 39}, 12707 (2006) [arXiv:0601049].\\
J. Hofmann,
"Dimensional reduction in quantum field theories at finite temperature and density",
Phys. Rev. D {\bf 82}, 125027 (2010) [arXiv:1009.4071].\\

\bibitem{NewInstability0}
F. Preis, A. Rebhan, A. Schmitt, "Inverse magnetic catalysis in dense holographic matter",
JHEP 1103 (2011) 033 [arXiv:1012.4785].\\  

\bibitem{NewInstability1}
S. A. Hartnoll, C. P. Herzog, G. T. Horowitz, "Holographic Superconductors",
JHEP 0812 (2008) 015 [arXiv:0810.1563].\\ 

\bibitem{NewInstability2}
N. Iqbal, H. Liu, M. Mezei, Q. Si, "Quantum phase transitions in holographic models of magnetism and superconductors",
Phys. Rev. D {\bf 82} 045002 (2010) [arXiv:1003.0010].\\ 

\bibitem{NewInstability3}
A. Adams, J. Wang, "Towards a Non-Relativistic Holographic Superfluid", 
New J. Phys. 13, 115008 (2011) [arXiv:1103.3472].\\

\bibitem{NewInstability4}
T. Faulkner, G. T. Horowitz, M. M. Roberts, "Holographic quantum criticality from multi-trace deformations",
JHEP 1104 (2011) 051 [arXiv:1008.1581].\\


\bibitem{Conductivity}
C. P. Herzog, P. Kovtun, S. Sachdev, D. T. Son, 
"Quantum critical transport, duality, and M-theory",
Phys. Rev. D {\bf 75}, 085020 (2007) [arXiv:0701036].\\ 
   
\bibitem{Conductivity2}
W. Witczak-Krempa, S. Sachdev,
"The quasi-normal modes of quantum criticallity",
Phys. Rev. B {\bf 86}, 235115 (2012) [arXiv:1210.4166].\\
R. C. Myers, S. Sachdev, A. Singh, 
"Holographic quantum critical transport without self-duality",
Phys. Rev. D {\bf 83}, 066017 (2011) [arXiv:1010.0443]. \\

\bibitem{rebhan}
A. Rebhan, A. Schmitt, and S. A. Stricker,
Meson supercurrents and the Meissner effect in
the Sakai- Sugimoto model,JHEP 05
(2009) 084, [arXiv:0811.3533]\\

\bibitem{sigmaDC1}
S. A. Hartnoll, C. P. Herzog, G. T. Horowitz, 
"Building an AdS/CFT superconductor",
[arXiv:0803.3295].\\ 

\bibitem{sigmaDC2}
A. Karch, A. O'Bannon, "Metallic AdS/CFT",
[arXiv:0705.3870].\\

\bibitem{sigmaDC3}
P. Basu, J. He, A. Mukherjee, H.-H. Shieh, 
"Superconductivity from D3/D7: Holographic Pion Superfluid",
[arXiv:0810.3970].\\

\bibitem{sigmaDC4}
C. P. Herzog, K.-W. Huang, R. Vaz, 
"Linear Resistivity from Non-Abelian Black Holes",
[arXiv:1405.3714].\\ 

\bibitem{p-waveTransport}
J. Erdmenger, D. Fernandez, H. Zeller, 
"New Transport Properties of Anisotropic Holographic Superfluids",
[arXiv:1212.4838].\\
P. Basu, J. He, A. Mukherjee, H.-H. Shieh, 
"Hard-gapped Holographic Superconductors",
[arXiv:0911.4999].\\ 

\bibitem{MetlitskiVishwanath}
M. A. Metlitski, A. Vishwanath, "Particle-vortex duality of 2D Dirac fermion from electric-magnetic duality
of 3D topological insulators", 
Phys.\ Rev.\ B {\bf 93}, 245151 (2016),
[arXiv:1505.05142].\\

\bibitem{WangSenthil}
C. Wang, T. Senthil, 
"Dual Dirac liquid on the surface of the electron topological insulator",
Phys.\ Rev.\ X {\bf 5}, 041031 (2015).\\
C. Wang, T. Senthil, 
"The half-filled Landau level, topological insulator surfaces, and three dimensional quantum spin liquids"
Phys.\ Rev.\ B {\bf 93}, 085110 (2016),
[arXiv:1507.08290].\\

\bibitem{MrossAliceaMotrunich}
D. F. Mross, J. Alicea, O. I. Motrunich,
"Explicit derivation of duality between a free Dirac cone and quantum electrodynamics in (2+1) dimensions",
Phys.\ Rev.\ Lett. {\bf 117}, 016802 (2016).\\

\bibitem{KarchTong}
A. Karch, D. Tong, 
"Particle-vortex duality from 3d bosonization",
Phys.\ Rev.\ X {\bf 6}, 031043 (2016),
[arXiv:1606.01893].\\
A. Karch, B. Robinson, D. Tong, "More abelian dualities in 2+1 dimensions",
[arXiv:1609.04012].\\

\bibitem{SeibergSenthilWangWitten}  
N. Seiberg, T. Senthil, C. Wang, E. Witten,
"A duality web in 2+1 dimensions and condensed matter physics",
[arXiv:1606.01989].\\
F. Benini, Po-Shen Hsin, N. Seiberg,
"Comments on global symmetries, anomalies, and duality in (2+1)d",
[arXiv:1702.07035].\\

\bibitem{vectormes1}
  M.~N.~Chernodub,
  ``Superconductivity of QCD vacuum in strong magnetic field,''
  Phys.\ Rev.\ D {\bf 82}, 085011 (2010)
  doi:10.1103/PhysRevD.82.085011
  [arXiv:1008.1055 [hep-ph]].
  
\bibitem{vectormes2}
  M.~N.~Chernodub,
  ``Electromagnetic superconductivity of vacuum induced by strong magnetic field,''
  Lect.\ Notes Phys.\  {\bf 871} (2013) 143,
  doi:10.1007/978-3-642-37305-3\_6
  [arXiv:1208.5025 [hep-ph]].
   
\bibitem{Aharony}
O. Aharony, 
"Baryons, monopoles and dualities in Chern-Simons-matter theories",
J. High Energy Phys. {\bf 1602}, 093 (2016),
[arXiv:1512.00161].\\

\bibitem{SakaiSugimoto}
  T.~Sakai and S.~Sugimoto,
  ``Low energy hadron physics in holographic QCD,''
  Prog.\ Theor.\ Phys.\  {\bf 113} (2005) 843
  doi:10.1143/PTP.113.843
  [hep-th/0412141].
    T.~Sakai and S.~Sugimoto,
  ``More on a holographic dual of QCD,''
  Prog.\ Theor.\ Phys.\  {\bf 114} (2005) 1083
  doi:10.1143/PTP.114.1083
  [hep-th/0507073].

\bibitem{Son}
D. T. Son, "The Dirac composite fermion of the fractional quantum Hall effect",
Prog. Theor. Exp. Phys. 12C, 103 (2016)
[arXiv:1608.05111].\\

\bibitem{aharony}
O. Aharony, K. Peeters, J. Sonnenschein, and M. Zamaklar,
"Rho meson condensation at finite
isospin chemical potential in a holographic model for QCD",
JHEP02 (2008) 071,
[arXiv:0709.3948].\\

\bibitem{MartinAmmon1}
M. Ammon, J. Erdmenger, M. Kaminski, and P. Kerner,
"Superconductivity from gauge/gravity duality with flavor",
Phys.\ Lett.\ B {\bf 680}, 516 (2009)
doi:	10.1016/j.physletb.2009.09.029
[arXiv:0810.2316 [hep-th]].\\
M. Ammon, J. Erdmenger, M. Kaminski, and P. Kerner,
"Flavor Superconductivity from Gauge/Gravity Duality",
JHEP 0910 (2009) 067
doi:	10.1088/1126-6708/2009/10/067
[arXiv:0903.1864 [hep-th]].\\

\bibitem{MartinAmmon2}
M. Ammon, J. Erdmenger, M. Kaminski, A. O'Bannon,
"Fermionic Operator Mixing in Holographic p-wave Superfluids",
JHEP 1005 (2010) 053
doi:	10.1007/JHEP05(2010)053
[arXiv:1003.1134 [hep-th]].\\

 \bibitem{kaminski}
 J. Erdmenger, M. Kaminski, P. Kerner, and F. Rust,
``Finite baryon and isospin chemical
potential in AdS/CFT with flavor'',JHEP
11(2008) 031 
[arXiv:0807.2663].\\
 M.~Kaminski,
 ``Flavor Superconductivity and Superfluidity,''
 Lect.\ Notes Phys.\  {\bf 828}, 349 (2011)
 [arXiv:1002.4886 [hep-th]].\\

\bibitem{stephanov} 
  D.~T.~Son and M.~A.~Stephanov,
  ``QCD and dimensional deconstruction,''
  Phys.\ Rev.\ D {\bf 69}, 065020 (2004)
  doi:10.1103/PhysRevD.69.065020
  [hep-ph/0304182].	\\
		
\bibitem{gk}
 A.~Gorsky and A.~Krikun,
  ``Baryon as dyonic instanton,''
  Phys.\ Rev.\ D {\bf 86}, 126005 (2012)
  doi:10.1103/PhysRevD.86.126005
  [arXiv:1206.4515 [hep-th]].\\
 A.~Gorsky, S.~B.~Gudnason and A.~Krikun,
  ``Baryon and chiral symmetry breaking in holographic QCD,''
  Phys.\ Rev.\ D {\bf 91}, no. 12, 126008 (2015)
  [arXiv:1503.04820 [hep-th]]. \\
	
\bibitem{karch}
	 M.~Fujita, M.~Kaminski and A.~Karch,
  ``SL(2,Z) Action on AdS/BCFT and Hall Conductivities,''
  JHEP 1207 (2012) 150
  doi:10.1007/JHEP07(2012)150
  [arXiv:1204.0012 [hep-th]].\\
  
\bibitem{burg2}
C. Burgess and B. P. Dolan, ``Particle Vortex Duality And The Modular Group:
Applications To The Quantum Hall Effect And Other 2-D Systems'', hep-th/0010246.\\

\bibitem{chernodub} 
  M.~N.~Chernodub,
  ``Superconductivity of QCD vacuum in strong magnetic field,''
 Phys.\ Rev.\ D {\bf 82}, 085011 (2010)
 doi:10.1103/PhysRevD.82.085011
 [arXiv:1008.1055 [hep-ph]].\\
	
\bibitem{Son:2007ny} 
 D.~T.~Son and M.~A.~Stephanov,
 ``Axial anomaly and magnetism of nuclear and quark matter,''
 Phys.\ Rev.\ D {\bf 77}, 014021 (2008)
 doi:10.1103/PhysRevD.77.014021
 [arXiv:0710.1084 [hep-ph]].\\

\bibitem{witten2} 
  E.~Witten,
  ``Baryons and branes in anti-de Sitter space,''
  JHEP 9807 (1998) 006,
  doi:10.1088/1126-6708/1998/07/006
  [hep-th/9805112].\\

\bibitem{orbi1} 
M. Hanada, Naoki Yamamoto, "Universality of Phases in QCD and QCD-like Theories", 
JHEP1202, 138 (2012
[arXiv:1103.5480].\\
 A.~Cherman and B.~C.~Tiburzi,
  ``Orbifold equivalence for finite density QCD and effective field theory,''
  JHEP 1106 (2011) 034,
  doi:10.1007/JHEP06(2011)034
  [arXiv:1103.1639 [hep-th]].\\

\bibitem{orbi2}
M. Hanada, C. Hoyos, A. Karch, L. G. Yaffe, 
"Holographic realization of large-Nc orbifold equivalence with non-zero chemical potential"
JHEP 1208 (2012) 081
[arXiv:1201.3718].\\

\bibitem{orbi3}
A.~Armoni and A.~Patella,
  ``QCD With A Chemical Potential, Topology, And The 't Hooft 1/N Expansion,''
  Phys.\ Rev.\ D {\bf 85}, 125021 (2012)
  doi:10.1103/PhysRevD.85.125021
  [arXiv:1204.2405 [hep-th]].\\

\bibitem{dolan:2006zc}
  B.~P.~dolan,
  ``Modular Symmetry and Fractional Charges in N=2 Supersymmetric Yang-Mills and the Quantum Hall Effect,''
  SIGMA {\bf 3} (2007) 010
  doi:10.3842/SIGMA.2007.010
  [hep-th/0611282].
  
  
  
\end{thebibliography}
\end{document}